\documentclass[11pt,a4paper]{article}

\usepackage[T1]{fontenc}
\usepackage[utf8]{inputenc}
\usepackage{multicol}
\usepackage{xspace}
\usepackage{centernot}
\usepackage{comment}
\usepackage{booktabs}
\usepackage{caption}
\usepackage{subcaption}
\usepackage{lineno}
\usepackage{mathrsfs}
\usepackage{xcolor}
\usepackage{mathtools} 
\usepackage{latexsym}
\usepackage{amssymb}
\usepackage{stmaryrd}
\usepackage{amsmath}
\usepackage{palatino}
\usepackage{authblk}
\usepackage{bm}

\SetSymbolFont{stmry}{bold}{U}{stmry}{m}{n}
\usepackage[hyperref,thmmarks]{ntheorem}

\theoremstyle{plain}
\theorembodyfont{\itshape}
\theoremsymbol{\ensuremath{\blacksquare}}

\newtheorem{prop}{Proposition}

\newtheorem{coro}{Corollary}
\newtheorem{fact}{Fact}

\theoremheaderfont{\normalfont\bfseries}
\theorembodyfont{\itshape}
\theoremsymbol{\ensuremath{\blacktriangleleft}}

\newtheorem{defi}{Definition}
\newtheorem{exam}{Example}
\newtheorem*{exam1}{Example 1, continued}

\theoremheaderfont{\itshape}
\theorembodyfont{\normalfont}
\theoremstyle{nonumberplain}
\theoremsymbol{\ensuremath{\blacksquare}}

\newtheorem{proof}{Proof.}

\usepackage[UKenglish]{babel}

\usepackage{natbib}

\setcitestyle{round,comma,authoryear,aysep={},yysep={,}}

\usepackage{hyperref}

\newcommand{\titulo}{Variations on distributed belief}
\newcommand{\autor}{John Lindqvist, Fernando R. Vel{\'a}zquez-Quesada, Thomas {\AA}gotnes}

\hypersetup{
  pdftitle          = {\titulo},
  pdfauthor         = {\textcopyright \autor},
  pdfstartview      = {FitH},
  pdfpagelayout     = {OneColumn},
  bookmarksnumbered = {true},
  naturalnames      = {true},
  hypertexnames     = false,
  colorlinks        = {true},           
}

\addto\extrasUKenglish{%
  \renewcommand{\sectionautorefname}{Section}%
  \renewcommand{\footnoteautorefname}{Footnote}
}

\newcommand{\autorefp}[1]{(\autoref{#1})}

\usepackage[textsize=footnotesize]{todonotes}

\definecolor{colorfer}{RGB}{128,0,255}

\definecolor{colorjohn}{RGB}{0,150,100}
\usepackage[inline]{enumitem}

\newcommand{\itmformat}{\bfseries \itshape}

\setlist[enumerate,1]{label={\itmformat (\arabic*)}}
\setlist[enumerate,2]{label={\itmformat (\alph*)}}

\newlist{compactenumerate}{enumerate}{1}
\setlist[compactenumerate,1]{label={\itmformat (\roman*)}, leftmargin=1.5em, rightmargin=0em, topsep=0.25em, itemsep=0.25em}

\newlist{inlineenum}{enumerate*}{1}
\setlist[inlineenum,1]{label={\itmformat (\roman*)}}

\setlist[itemize,1]{label=\textbullet}
\setlist[itemize,2]{label=--}

\newlist{compactitemize}{itemize}{3}
\setlist[compactitemize,1]{label=\textbullet, leftmargin=1.5em, rightmargin=0em, topsep=0.25em, itemsep=0.25em}
\setlist[compactitemize,2]{label=--, leftmargin=1.25em, rightmargin=0em, topsep=0.25em, itemsep=0.25em}
\setlist[compactitemize,3]{label=$\cdot$, leftmargin=0.25em, rightmargin=0em, topsep=0.25em, itemsep=0.25em}

\newlist{bisimitemize}{itemize}{1}
\setlist[bisimitemize,1]{label=, leftmargin=0.75em, rightmargin=0em, topsep=0.25em, itemsep=0.25em}

{ \normalcolor }

\newcommand{\hideable}[1]{}

\newcommand{\gcsbasic}[1]{C_{#1}}            
\newcommand{\gcs}[2]{\gcsbasic{#1}(#2)}      
\newcommand{\cgcs}[2]{C^{\forall}_{#1}(#2)}  
\newcommand{\crel}[1]{R^\forall_{#1}}        

\newcommand{\maxatin}[3]{#1 \subseteq^{max}_{#2} #3}
\newcommand{\maxat}[2]{\mathit{MC}_{#1}(#2)} 
\newcommand{\maxcsat}[2]{\mathit{CC}_{#1}(#2)} 

\newcommand{\colleadsto}[1]{\mathrel{\leadsto_{#1}}}
\newcommand{\colmcssleadto}[1]{\mathrel{\leadsto^{max}_{#1}}}

\newcommand{\Dia}{\Diamond}

\newcommand{\msf}{\mathscr{F}}

\newcommand{\modalityB}{\operatorname{B}}
\newcommand{\beli}[2]{\mathop{\modalityB_{#1}}#2}

\newcommand{\modalityD}{\operatorname{D}}
\newcommand{\dstan}[2]{\mathop{\modalityD_{#1}}#2}
\newcommand{\ddstan}[2]{\mathop{\widehat{\modalityD}_{#1}}#2}
\newcommand{\dcaut}[2]{\mathop{\modalityD^{\forall}_{#1}}#2}

\newcommand{\dbold}[2]{\mathop{\modalityD^{\exists}_{#1}}#2}
\newcommand{\ddbold}[2]{\mathop{\overbracket[0.6pt][1pt]{\modalityD}\,\!^{\exists}_{#1}}#2}

\newcommand{\eqdcaut}{\mathrel{\leftrightsquigarrow_{\dcaut{}{}}}}

\newcommand{\eqdbold}{\mathrel{\leftrightsquigarrow_{\dbold{}{}}}}

\newcommand{\bidcautover}[1]{\leftrightarroweq^{#1}_{\dcaut{}{}}}
\newcommand{\bidcaut}{\bidcautover{}}
\newcommand{\bidboldover}[1]{\leftrightarroweq^{#1}_{\dbold{}{}}}
\newcommand{\bidbold}{\bidboldover{}}

\newcommand{\lang}[1]{\ensuremath{\mathcal{L}_{#1}}\xspace}
\newcommand{\limp}{\rightarrow}
\newcommand{\ldimp}{\leftrightarrow}

\newcommand{\cond}{\mathcal{F}}
\newcommand{\inc}[1]{{\asymp_{#1}}}

\newcommand{\M}{\mathcal{M}}
\newcommand{\dom}{\operatorname{D}}
\newcommand{\val}{V}
\newcommand{\mps}[1]{\M,#1 \vDash}
\newcommand{\mpf}[1]{\M,#1 \nvDash}

\newcommand{\class}[1]{\textnormal{\textbf{#1}}} 

\newcommand{\bdot}{\,\bm{.}\,}
\newcommand{\set}[1]{\{#1\}}
\newcommand{\card}[1]{\lvert #1 \rvert}
\newcommand{\comp}[1]{\overline{#1}}
\newcommand{\extm}[1]{\llbracket #1\rrbracket_{\M}}

\newcommand{\prooflr}{\ensuremath{\bm{(\Rightarrow)}}\xspace}
\newcommand{\proofrl}{\ensuremath{\bm{(\Leftarrow)}}\xspace}

\renewcommand{\emptyset}{\varnothing}

\newcommand{\msparagraph}[1]{\medskip\noindent\textbf{#1}\,}     

\newenvironment{ctabular}[1]
{\begin{center}\begin{tabular}{#1}}
{\end{tabular} \end{center}}

\newenvironment{smallctabular}[1]
{\begin{center}\begin{small}\begin{tabular}{#1}}
{\end{tabular}\end{small}\end{center}}

\newenvironment{footnotesizectabular}[1]
{\begin{center}\begin{footnotesize}\begin{tabular}{#1}}
{\end{tabular}\end{footnotesize}\end{center}}

\newenvironment{keyword}
{\noindent\textbf{Keywords:}\xspace}
{}

\newcommand{\sep}{\ensuremath{\cdot}\;}

\usepackage{tikz}
\usetikzlibrary{positioning,arrows,calc,fit,backgrounds,automata,shapes}

\colorlet{verylightgray}{lightgray!75}

\tikzset{
  every picture/.style = {
    thick,
    ->,
    >=stealth',
  }
  ,
  within/.style = {
    fill = white,
    inner sep = 1pt
  }
  ,
  frame rectangle/.style = {
    framed,
    draw=black,
    rounded corners,
  }
  ,
  math mode/.style = {
    execute at begin node=$,
    execute at end node=$
  }
  ,
  modality/.style = {
    minimum size=12pt,
    math mode,
    font = \small,
  }
  ,
  world/.style = {
    rectangle,
    rounded corners = 5,
    draw,
    minimum size=12pt,
    math mode,
    fill=gray!15,
    font = \footnotesize,
  }
  ,
  label edge/.style = {
    font = \footnotesize,
  }
  ,
  bisim/.style = {
    dashed,
    -
  }
  ,
  point/.style = {
    circle,
    draw,
    inner sep=0.5mm,
    fill=black
  }
  ,
  reflexive above/.style = {
    ->,
    loop,
    looseness=7,
    in=120,
    out=60
  }
  ,
  reflexive below/.style = {
    ->,
    loop,
    looseness=7,
    in=240,
    out=300
  }
  ,
  reflexive left/.style = {
    ->,
    loop,
    looseness=7,
    in=150,
    out=210
  }
  ,
  reflexive right/.style = {
    ->,
    loop,
    looseness=7,
    in=30,
    out=330
  }
}

\newcommand{\caut}{cautious\xspace}

\begin{document}

\renewcommand{\sectionautorefname}{Section}%
\renewcommand{\footnoteautorefname}{Footnote}


\title{\titulo}

\author[1]{John Lindqvist}
\author[2]{Fernando R. Vel{\'a}zquez-Quesada}
\author[3]{Thomas {\AA}gotnes}

\affil[1,2,3]{Department of Information Science and Media Studies, Universitetet i Bergen\\%
\texttt{\{John.Lindqvist,Fernando.VelazquezQuesada,Thomas.Agotnes\}@uib.no}%
}
\affil[3]{Institute of Logic and Intelligence, Southwest University}

\date{}

\maketitle


\begin{abstract}
  Motivated by the search for forms of distributed belief that do not collapse in the face of conflicting information, this paper introduces the notions of \emph{cautious} and \emph{bold} distributed belief. Both notions rely on maximally consistent subgroups of agents, with \emph{cautious} quantifying universally and \emph{bold} quantifying existentially. As a result, while the cautious distributed belief of a group is inconsistent only when
  all group members are individually inconsistent, the bold distributed belief of a group is never inconsistent. The paper discusses these two notions, presenting their respective modalities and semantic interpretations, discussing some of their basic properties, studying whether they preserve doxastic properties from the members of the group, and comparing them not only with standard distributed belief but also with one another, both at the level of modalities and at the level of language expressivity.
\end{abstract}


\section{Introduction}

Epistemic logic (\textit{EL}; \citealp{Hintikka1962}) is a simple and yet powerful framework for representing and reasoning about the knowledge of a set of agents. It typically relies on relational \emph{Kripke} models, assigning to each agent a binary \emph{accessibility} relation over possible worlds (i.e., possible states of affairs). Each relation represents the corresponding agent's \emph{uncertainty}, which is then used to define her \emph{knowledge}: at a world $w$ an agent $i$ knows that $\varphi$ is the case if and only if $\varphi$ holds in all the situations that are, for her, accessible from $w$. Despite its simplicity, \textit{EL} has become a widespread tool, contributing to the formal study of complex multi-agent epistemic phenomena in philosophy \citep{Hendricks2006}, computer science \citep{FaginEtAl1995,MeyervanDerHoek1995elaics} and economics \citep{deBruin2010,Perea2012}.

One of the most attractive features of \textit{EL} is that one can reason not only about individual knowledge, but also about different forms of knowledge for groups. A historically important example is the notion of \emph{common knowledge} \citep{Lewis1969}, known to be crucial in social interaction.\footnote{A group has common knowledge of $\varphi$ if and only if everybody in the group knows $\varphi$, everybody in the group knows that everybody in the group knows $\varphi$, and so on.} Another important epistemic notion for groups, key in distributed systems, is that of \emph{distributed knowledge} \citep{Hilpinen1977,HalpernMoses1985,HalpernMoses1990}. Intuitively, a group has distributed knowledge of $\varphi$ if and only if $\varphi$ follows from the combination of the individual knowledge of all its members. In \textit{EL}, which uses uncertainty to define knowledge, this intuition has a natural representation: at a world $w$ a group $G$ has distributed knowledge of $\varphi$ if and only if $\varphi$ holds in all the situations that \emph{all the members of the group} can reach (i.e., in all the situations \emph{no one in the group} can distinguish) from $w$. Thus, the accessibility relation for the distributed knowledge of a group $G$ corresponds to the \emph{intersection} of the accessibility relations of $G$'s members.

\smallskip

Since distributed knowledge is the result of combining the individual knowledge of different agents, one can wonder whether agents might have contradictory distributed knowledge (i.e., whether it is possible for a set of agents to know a contradiction distributively, which semantically corresponds to the intersection of the agents' relations being empty). When one works with a \emph{truthful} notion of knowledge (semantically, when all accessibility relations are required to be reflexive), distributed knowledge does not have this problem: all relations contain the reflexive edges, and thus their intersection will never be empty. However, when one works with weaker notions of information, contradictions might arise. For example, if one works with a notion of belief represented by a serial, transitive and Euclidean relation (see, e.g., \citealp{Hintikka1962}), it is possible for all agents to be consistent (i.e., none of them believes contradictions), and yet their distributed beliefs might be inconsistent -- even if it is possible to combine the information of \emph{some} of them in a consistent way.

\smallskip

This paper studies two alternative forms of (group) distributed belief, \emph{cautious} and \emph{bold} (modalities $\dcaut{}$ and $\dbold{}$, respectively), which do not collapse in the face of conflicting information.
Indeed, neither of these notions need to become inconsistent when the whole group is \emph{collectively} inconsistent, picking out instead a form of maximally consistent combined information. The intuition behind it is that, although a group $G$ as a whole might be inconsistent at some world $w$ (i.e., the set of worlds everyone in the group can access from $w$ is empty), there might be consistent subgroups. Among them, the maximally consistent groups are particularly relevant, as they represent the different ways of consistently combining information from as many agents as possible. Considering notions of maximal consistency is a standard approach in the literature (in particular, in non-monotonic reasoning) for resolving potential conflicts. Think, e.g., about the \emph{extensions} of a theory in default logic \citep{Reiter80}, or the \emph{maximally admissible} (i.e., preferred) sets of arguments in abstract argumentation theory \citep{Dung1995}. The idea has been also used within epistemic logic (e.g., by \citealp{vanBenthemPacuit2011} in the context of evidence-based beliefs) and also for distributed beliefs (by \citealp{herzig2020}, in the context of \emph{explicit} beliefs defined via belief bases).
 
The definitions of both cautious and bold distributed belief rely on these maximally consistent subgroups of agents. Yet, they use them in different ways. On the one hand, the first follows a cautious approach: at a world $w$ a group $G$ has cautious distributed belief that $\varphi$ if and only if \emph{every} maximally consistent subgroup of $G$ has distributed belief that $\varphi$. This corresponds to the \emph{skeptical} reasoner in non-monotonic reasoning. On the other hand, the second follows a bold approach: at a world $w$ a group $G$ has bold distributed belief that $\varphi$ if and only if \emph{some} maximally consistent subgroup of $G$ has distributed belief that $\varphi$. This corresponds to the \emph{credulous} reasoner in non-monotonic reasoning.

\msparagraph{Outline} The paper is organised as follows. \autoref{sec:basic.def} recalls the definition of a relational (Kripke) model (called here belief model) as well as that of the standard modal operators for individual belief ($\beli{}$) and distributed belief ($\dstan{}$). It also recalls the basic tools for comparing language expressivity, as they will be used throughout the text to contrast languages based on the discussed modalities. \autoref{sec:DA} focusses on \emph{cautious} distributed belief, introducing the modality and its semantic interpretation, discussing some of its basic properties, studying whether it preserves properties of the individual beliefs of the group's members, and comparing it with standard distributed belief, both at the level of modalities and at the level of language expressivity. While these results have been already presented in the conference paper \citet{LindqvistVA22} (of which this text is an extension), the presentation here is more streamlined.

\autoref{sec:DE} contains the novel contribution of this paper: the study of \emph{bold} distributed belief. It follows the structure of the section that precedes it, introducing the modality and its semantic interpretation, discussing some of its basic properties, studying the preservation of doxastic properties from individuals to a group, and comparing it (again, at the level of modalities and at the level of language expressivity), this time not only with standard distributed belief but also with cautious distributed belief. Finally, \autoref{sec:end} summarises the results, discussing also further research lines.

\section{Basic definitions}\label{sec:basic.def}

Throughout this text, let $A$ be a finite non-empty set of agents and $P$ be a countable non-empty set of atomic propositions. The subsets of $A$ discussed in this text, denoted by $G$, $H$ and so on, will be \emph{non-empty}. In other words, a \emph{group} of agents $G$ is a \emph{non-empty} subset of $A$, and a \emph{subgroup} of $G$ is a \emph{non-empty} subset of the group $G$.
For languages, the propositional one (with $\lnot$ and $\land$ as primitive operators) is denoted by \lang{} (its semantic interpretation being the usual). Then, $\lang{\mathit{X1}, \dots, \mathit{Xn}}$ is the language that extends \lang{} with the modalities $\mathit{X1}, \dots, \mathit{Xn}$. When a modality $X$ uses a subindex $\iota$ (with $\iota$ an agent or a group of them), including it in a language means to include all modalities $X_\iota$.\footnote{In particular, while the language $\lang{\dstan{}{}}$ is \lang{} plus a unary modality $\dstan{G}$ for each group $\emptyset \subset G \subseteq A$, the language $\lang{\dcaut{}{}}$ is \lang{} plus a unary modality $\dcaut{G}$ for each $\emptyset \subset G \subseteq A$ and the language $\lang{\dbold{}{}}$ is \lang{} plus a unary modality $\dbold{G}$ for each $\emptyset \subset G \subseteq A$.}

\smallskip

Then, the model.

\begin{defi}[Belief model]\label{def:model}
  A \emph{belief model} is a tuple $\M = \langle W,R,\val \rangle$ in which $W$ (also denoted as $\dom(\M)$) is a non-empty set of objects called possible worlds, $R = \set{R_a \subseteq W \times W \mid a \in A}$ assigns an arbitrary \emph{accessibility} relation to each agent $a \in A$,\footnote{We use infix notation for relations indexed by (sets of) agents.} and $\val:P \rightarrow 2^W$ is a valuation function. A \emph{pointed belief model} is a pair $(\M, w)$ with $\M$ a belief model and $w \in \dom(\M)$ its evaluation point. The class of all belief models is denoted as \class{M}. Given $\langle W,R,\val \rangle$ in \class{M}, $a \in A$ and $w \in W$, the set $\gcs{a}{w} := \set{w' \in W \mid w R_a w'}$ is called \emph{$a$'s conjecture set at $w$}. One can generalise $R_a$ and $\gcsbasic{a}$ to a group of agents $G \subseteq A$ by defining $R_G := \bigcap_{a \in G} R_a$ (the distributed accessibility relation for $G$) and $\gcs{G}{w} := \bigcap_{a \in G} \gcs{a}{w}$ (\emph{$G$'s (combined) conjecture set at $w$}, equivalently defined as $\gcs{G}{w} := \set{w' \in W \mid w R_G w'}$).
\end{defi}

Belief models are nothing but multi-agent Kripke (relational) models in which the relations are arbitrary (thus, the term ``belief'' is used in a rather loose way). They allow us to represent not only the beliefs each individual agent has, but also different forms of belief for groups. This can be done in different ways, but a common strategy follows two steps. First, use the accessibility relations of the members of a group $G$ to define an accessibility relation for some form of belief for the group. Then, use this accessibility relation in the `standard way' (a universal/existential quantification over the worlds that can be reached via the relation) to define the semantic interpretation of a modality representing this form of group belief. A modality defined in this way is said to have \emph{relational semantics}.

\smallskip

The semantic interpretation of the group modality for standard distributed beliefs, $\dstan{}$, together with that of the standard operator for individual belief, $\beli{}$, is recalled below. Note: $\dstan{}$ has relational semantics, as it is semantically interpreted as a universal quantification over the worlds that can be reached via the relation $R_G$ (for the given group $G \subseteq A$), with $R_G$ defined as the intersection of the relations of the group's members.

\begin{defi}[Distributed belief]\label{def:D}
  Let $(\M, w)$ be a pointed belief model with $\M = \langle W,R,\val \rangle$; take $a \in A$ and group $G \subseteq A$. Then,
  \begin{center}
    \begin{tabular}[b]{l@{\qquad{iff}\qquad}l}
      $\M,w \vDash \beli{a}{\varphi}$  & $\forall w' \in W$: if $wR_aw'$ then $\M,w' \vDash \varphi$, \\
                                       & $\forall w' \in \gcs{a}{w}$: $\M,w' \vDash \varphi$, \\
      $\M,w \vDash \dstan{G}{\varphi}$ & $\forall w' \in W$: if $wR_Gw'$ then $\M,w' \vDash \varphi$, \\
                                       & $\forall w' \in \gcs{G}{w}$: $\M,w' \vDash \varphi$, \\
    \end{tabular}
  \end{center}
  The set $\extm{\varphi} := \set{s \in W \mid \M,s \vDash \varphi}$ contains all worlds in $\M$ where $\varphi$ holds. A formula $\varphi$ is \emph{valid in a class of belief models \class{C}} (notation: $\class{C} \vDash \varphi$) when $\varphi$ is true in every world of every model in \class{C}. The formula is \emph{valid} (notation: $\vDash \varphi$) when $\class{M} \vDash \varphi$.
\end{defi}

Observe how individual belief operators $\beli{a}{}$ can be defined in terms of $\dstan{}$, as $\dstan{\set{a}}{\varphi}$ (abbreviated as $\dstan{a}{\varphi}$) holds in a world $w$ if and only if $\M,w' \vDash \varphi$ holds for all $w' \in \gcs{a}{w}$. The dual of $\dstan{}{}$ is defined in the standard way, $\ddstan{G}{\varphi} := \lnot \dstan{G}{\lnot \varphi}$, so
\begin{center}
  \begin{tabular}[b]{l@{\qquad{iff}\qquad}l}
    $\M,s \vDash \ddstan{G}{\varphi}$ & $\exists s' \in \gcs{G}{s}$: $\M,s' \vDash \varphi$. \\
  \end{tabular}
\end{center}

\medskip

For comparing the expressivity of different languages, the following definition will be useful.

\begin{defi}[Relative expressivity]\label{def:expressivity}
  Let $\lang{1}$ and $\lang{2}$ be two languages whose formulas can be evaluated over pointed belief models.
  \begin{compactitemize}
    \item $\lang{2}$ is \emph{at least as expressive} as $\lang{1}$ (notation: $\lang{1} \preccurlyeq \lang{2}$) if and only if every formula in $\lang{1}$ has a semantically equivalent formula in $\lang{2}$: for every $\alpha_1 \in \lang{1}$ there is $\alpha_2 \in \lang{2}$ s.t., for every pointed belief model $(\M, w)$, we have $\mps{w} \alpha_1$ if and only if $\mps{w} \alpha_2$.

    \item $\lang{1}$ and $\lang{1}$ are \emph{equally expressive} (notation: $\lang{1} \approx \lang{2}$) if and only if $\lang{1} \preccurlyeq \lang{2}$ and $\lang{2} \preccurlyeq \lang{1}$.

    \item $\lang{2}$ is \emph{strictly more expressive} than $\lang{1}$ (notation: $\lang{1} \prec \lang{2}$) if and only if $\lang{1} \preccurlyeq \lang{2}$ and $\lang{2} \not\preccurlyeq \lang{1}$.
  \end{compactitemize}
\end{defi}

Note: a typical strategy for proving $\lang{1} \preccurlyeq \lang{2}$ is to give a translation $\mathit{tr}:\lang{1} \to \lang{2}$ such that for every $\alpha_1 \in \lang{1}$ and every $(\M, w)$ we have $\mps{w} \alpha_1$ if and only if $\mps{w} \mathit{tr}(\alpha_1)$. The crucial cases are those for the operators in $\lang{1}$ that do not occur in $\lang{2}$. Note also: a typical strategy for proving $\lang{1} \not\preccurlyeq \lang{2}$ is to find two pointed models that satisfy exactly the same formulas in $\lang{2}$, and yet can be distinguished by a formula in $\lang{1}$.

\section{Cautious distributed belief}\label{sec:DA}

\subsection{Semantic interpretation}

This section focusses on the novel notion of \emph{\caut distributed belief} (modality: $\dcaut{}$), together with its relationship with \emph{distributed belief}. For the semantic interpretation of the first, the following definitions will be useful.

\begin{defi}[Consistency, maximal consistency]\label{def:consistency}
  Take $\langle W,R,\val \rangle$ in \class{M}. Take groups of agents $G' \subseteq G \subseteq A$ and a world $w \in W$. The group of agents $G'$ is \emph{consistent at $w$} if and only if $\gcs{G'}{w} \neq \emptyset$. It is \emph{maximally consistent at $w$ w.r.t. $G$} (notation: $\maxatin{G'}{w}{G}$) if and only if it is consistent at $w$ and, additionally, every group $H$ satisfying $G' \subset H \subseteq G$ is such that $\gcs{H}{w} = \emptyset$ (i.e., every strict superset of $G'$ up to $G$ is inconsistent). Finally, the set $\cgcs{G}{w} := \bigcup_{\maxatin{G'}{w}{G}} \gcs{G'}{w}$ (the \emph{consistent (combined) conjecture set} of $G$ at $w$) contains the worlds that are relevant for the maximally consistent subgroups of $G$ at world $w$. The \emph{relation} for cautious distributed belief $\crel{G} \subseteq \dom(\M) \times \dom(\M)$, defined as $w \crel{G} w'$ iff $w' \in \cgcs{G}{w}$, will simplify some later work.
\end{defi}

Here is the semantic interpretation for $\dcaut{}$. We also present the semantics of an additional constant $\inc{G}$, which will be useful later.


\begin{defi}[Cautious distributed belief]\label{def:DA}
  Let $(\M, w)$ be a pointed belief model with $\M = \langle W,R,\val \rangle$; take group $G \subseteq A$. Then,
  \begin{center}
    \begin{tabular}[b]{l@{\qquad{iff}\qquad}l}
      $\M,w \vDash \dcaut{G}{\varphi}$ & $\forall \maxatin{G'}{w}{G}$: $\M,w \vDash \dstan{G'}{\varphi}$ \\
      $\M,w \vDash \inc{G}$            & $\gcs{G}{w} = \emptyset$.
    \end{tabular}
  \end{center}
  Validity (in a class of belief models) is defined and denoted as before.
\end{defi}

As mentioned in the introduction, $\dcaut{G}{\varphi}$ holds at $w$ if and only if \emph{every} maximally consistent subgroup of $G$ has, at $w$, distributed belief that $\varphi$. Still, expanding the definition of distributed belief in this clause provides a better look at what cautious distributed belief is:
\begin{center}
  \begin{tabular}[b]{l@{\qquad{iff}\qquad}l}
    $\M,w \vDash \dcaut{G}{\varphi}$ & $\forall \maxatin{G'}{w}{G}$, $\forall w' \in \gcs{G'}{w}$: $\M,w' \vDash \varphi$. \\
  \end{tabular}
\end{center}
Note now the difference between $\dstan{G}{}$ and $\dcaut{G}{}$. On the one hand, $\dstan{G}{\varphi}$ holds at $w$ when every world in the conjecture set of \emph{$G$} at $w$ satisfies $\varphi$. On the other hand, $\dcaut{G}{\varphi}$ holds at $w$ when every world $w'$ in the conjecture set at $w$ of \emph{every subgroup of $G$ that is maximally consistent at $w$} satisfies $\varphi$. This alternative version of the definition makes explicit the two universal quantifications: first over maximally consistent subgroups and then over the worlds in these groups' conjecture sets.

Another equivalent definition sheds additional light on cautious distributed belief. By using the relation for cautious distributed belief of \autoref{def:consistency} (equivalently, the group's  consistent conjecture set), one gets
\begin{center}
  \phantomsection
  \label{cautious.normal}
  \begin{tabular}[b]{l@{\qquad{iff}\qquad}l}
    $\M,w \vDash \dcaut{G}{\varphi}$ & $\forall w' \in W$: if $w \crel{G}w'$ then $\M,w' \vDash \varphi$, \\
                                     & $\forall w' \in \cgcs{G}{w}$: $\M,w' \vDash \varphi$.
  \end{tabular}
\end{center}
This shows that $\dcaut{G}$ has relational semantics and thus it is, in fact, a normal modality. Therefore, e.g., it distributes over implications ($\vDash \dcaut{G}{(\varphi \limp \psi)} \limp (\dcaut{G}{\varphi} \limp \dcaut{G}{\psi})$) and `contains' all validities (if $\vDash \varphi$ then $\vDash \dcaut{G}{\varphi}$).

\smallskip

Finally, note how $\inc{G}$ simply expresses the fact that the group of agents $G$ is inconsistent at the evaluation point.

\medskip

Here is a simple example of the use of $\dcaut{}$ for describing a belief model, comparing it with the standard $\dstan{}$.

\begin{exam}\label{exa:DA:1}
  Consider the pointed belief model $(\M, w_1)$ below, with the evaluation point double circled (the individual relations are serial, transitive and Euclidean, the properties most commonly associated with a belief operator). In it (i.e., at $w_1$), $a$ believes $p$ to be true and $q$ to be false ($\M,w_1, \vDash \beli{a}{p} \land \beli{a}{\neg q}$). Nevertheless, $b$ is uncertain about $p$ but believes $q$ to be true ($\M,w_1 \vDash (\neg \beli{b}{p} \land \neg \beli{b}{\neg p}) \land \beli{b}{q}$). Finally, $c$ believes $p$ but is uncertain about $q$ ($\M,w_1, \vDash \beli{c}{p} \land (\neg \beli{c}{q} \land \neg \beli{c}{\neg q})$).
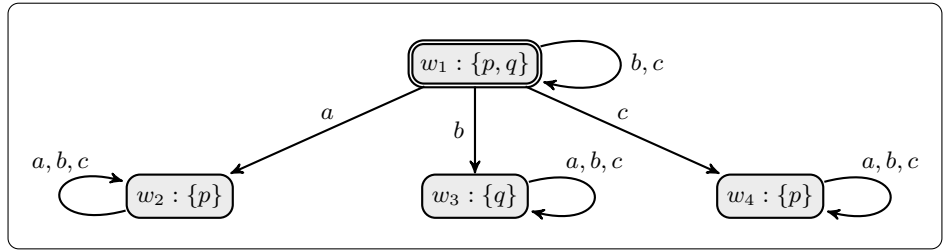
  \begin{center}
    \begin{tikzpicture}[frame rectangle]
      \node[world, double] (w) {w_1:\set{p,q}};
      \node[world, below left = 3em and 6em of w] (u1) {w_2:\set{p}};
      \node[world, below = 3em of w] (u2) {w_3:\set{q}};
      \node[world, below right = 3em and 6em of w] (u3) {w_4:\set{p}};

      \path (w)  edge node [above, label edge] {$a$} (u1)
                 edge node [left, label edge] {$b$} (u2)
                 edge [loop right] node [right, label edge] {$b,c$} ()
                 edge node[above, label edge] {$c$} (u3)
            (u1) edge [loop left] node [above = 5pt, label edge] {$a,b,c$} ()
            (u2) edge [loop right] node [above = 5pt, label edge] {$a,b,c$} ()
            (u3) edge [loop right] node [above = 5pt, label edge] {$a,b,c$} ();
    \end{tikzpicture}
  \end{center}
  Consider first the group $G_1=\set{a,b}$. On the one hand, both members of $G_1$ are individually consistent at $w_1$ and yet $\gcs{G_1}{w_1}=\emptyset$; thus, at $w_1$, the maximally consistent subgroups are $\set{a}$ and $\set{b}$. Their conjecture sets are $\gcs{a}{w_1}=\set{w_2}$ and $\gcs{b}{w_1}=\set{w_1,w_3}$, and hence $G_1$'s consistent conjecture set is $\cgcs{G_1}{w_1}=\set{w_1,w_2,w_3}$. Thus, $\M,w_1 \vDash \neg \dcaut{G_1}{p} \land \neg \dcaut{G_1}{q}$. On the other hand, when we consider standard distributed belief, we see that $\M,w_1 \vDash \dstan{G_1}{p} \land \dstan{G_1}{q}$. This is due to the fact that $\gcs{G_1}{w_1}=\emptyset$ and we end up quantifying over an empty set. Thus, we also get $\M,w_1 \vDash \dstan{G_1}{\bot}$.

  Now $c$ joins the group, $G_2 = \set{a,b,c}$. On the one hand, $b$ and $c$ are consistent at $w_1$ (i.e., they can `consistently combine information'); still, $a$ and $c$ are not (and, as mentioned before, neither are $a$ and $b$). Thus, the maximally consistent sets are $\set{a}$ and $\set{b,c}$. The relevant conjecture sets are now $\gcs{a}{w_1}=\set{w_2}$ and $\gcs{\set{b,c}}{w_1}=\set{w_1}$, so $\cgcs{G_2}{w_1}= \set{w_1,w_2}$. Then, $\M,w_1 \vDash \dcaut{G_2}{p} \land \neg \dcaut{G_2}{q}$ (the latter because, even though $b$ and $c$ together believe $q$, agent $a$ remains `a loner' and still believes that $q$ is false). On the other hand, the situation with standard distributed belief remains as for $G_1$: $\M,w_1 \vDash \dstan{G_2}{p} \land \dstan{G_2}{q} \land \dstan{G_2}{\bot}$.
\end{exam}

\subsection{Some basic results about \texorpdfstring{$\dcaut{}$}{DA}}\label{sub:DA:basics}

The proposition below presents three basic properties of the cautious distributed belief modality. The first one characterises the situations in which a group has cautious distributed belief in contradictions, i.e., the situations in which this form of group belief is inconsistent.
Indeed, while distributed beliefs for a group $G$ can be inconsistent (even when every agent in $G$ is consistent), for cautious distributed beliefs this is not the case: they are inconsistent if and only if \emph{all} agents in $G$ are inconsistent. The second tells us that individual belief operators ($\beli{a}{}$ for $a \in A$) can be expressed using $\dcaut{\set{a}}$, just as with the modality for standard distributed belief. Finally, an important property of standard distributed belief is \emph{group monotonicity}: if a group $G \subseteq A$ has standard distributed belief that $\varphi$, then so does any extension $H \supseteq G$ (thus, $G \subseteq H \subseteq A$ implies $\vDash \dstan{G}{\varphi} \rightarrow \dstan{H}{\varphi}$). This fails for $\dcaut{}$.

\begin{prop}\label{pro:DA}
  Take any groups $G \subseteq H \subseteq A$ and any $a \in A$.
  \begin{multicols}{2}
    \begin{enumerate}
      \item\label{pro:DA:itm:consistency} $\displaystyle \vDash \dcaut{G}{\bot} \leftrightarrow \bigwedge_{a \in G} \beli{a}{\bot}$.

      \item\label{pro:DA:itm:singleton} $\displaystyle \vDash \beli{a}{\varphi} \leftrightarrow \dcaut{\set{a}}{\varphi}$.

      \item\label{pro:DA:itm:coalition-mono} $\displaystyle \nvDash \dcaut{G}{\varphi} \rightarrow \dcaut{H}{\varphi}$.
    \end{enumerate}
  \end{multicols}
  \begin{proof}
    Take any $\M$, any $w \in \dom(\M)$.
    \begin{enumerate}
      \item $\prooflr$ If $\mps{w} \dcaut{G}{\bot}$ then, because no world satisfies $\bot$, either $\gcs{G'}{w} = \emptyset$ for all $\maxatin{G'}{w}{G}$, or there is no $G'$ satisfying $\maxatin{G'}{w}{G}$. But, by definition, no $G'$ satisfying $\maxatin{G'}{w}{G}$ is s.t. $\gcs{G'}{w} = \emptyset$ (i.e., no maximally consistent set is inconsistent). Hence, there is no $G'$ satisfying $\maxatin{G'}{w}{G}$, which means every $G' \subseteq G$ is s.t. $\gcs{G'}{w} = \emptyset$. In particular, all singletons $\set{a}$ for $a \in G$ are s.t. $\gcs{a}{w} = \emptyset$, and thus $\mps{w} \bigwedge_{a \in G} \beli{a}{\bot}$. $\proofrl$ If $\mps{w} \bigwedge_{a \in G} \beli{a}{\bot}$ then $\gcs{a}{w} = \emptyset$ for every $a \in G$. Hence, every non-empty $G' \subseteq G$ is s.t. $\gcs{G'}{w} = \emptyset$, so there is no $G'$ satisfying $\maxatin{G'}{w}{G}$. Thus, $\mps{w} \dcaut{G}{\bot}$.

      \item Suppose $\gcs{a}{w}= \emptyset$. Then both $\mps{w} \beli{a}{\varphi}$ and $\mps{w} \dcaut{a}{\varphi}$ hold vacuously for any $\varphi$. Otherwise, $\gcs{a}{w}\neq \emptyset$ and then $\set{a}$ is the only maximally consistent subset of $\set{a}$. Thus, $\mps{w}\dcaut{a}{\varphi}$ iff $\forall w' \in \gcs{a}{w}$: $\mps{w}\varphi$ iff $\mps{w}B_a\varphi$.

      \item Suppose $\mps{w} \dcaut{G}{\varphi}$ and consider a set $H \supseteq G$. In general, agents in $H \setminus G$  might not be consistent with any agent in $G$. In such cases, when consistent, these new agents will be part of a different maximally consistent subgroup, which might not have the distributed belief $\varphi$. This is shown in \autoref{exa:DA:1}, where $\M,w_1 \vDash \dcaut{\set{b}}{q}$ and yet $\M,w_1 \nvDash \dcaut{\set{a,b}}{q}$.
    \end{enumerate}
  \end{proof}
\end{prop}

\subsection{Preserving doxastic properties}\label{sub:DA:preserving}

As stated earlier, this paper has used the term ``belief'' in a loose way. Indeed, although the doxastic notion represented by the single-agent modality $\beli{}{}$ has the standard properties of a normal modality (it distributes over implications and it `contains' all validities), it does need to satisfy additional properties one might associate with a proper notion of belief. In particular, $\beli{}{}$ does not need to be consistent ($\beli{a}{\varphi} \limp \lnot \beli{a}{\lnot \varphi}$ is not valid), and it needs to be neither positively nor negatively introspective (neither $\beli{a}{\varphi} \limp \beli{a}{\beli{a}{\varphi}}$ nor $\lnot \beli{a}{\varphi} \limp \beli{a}{\lnot \beli{a}{\varphi}}$ are valid). Still, given $\beli{}{}$'s relational semantics and the well-known correspondence between different relational properties and the validity of certain formulas \citep[Chapter 3]{Blackburn2002}, it is straightforward to obtain such properties for the modalities.
Using $S$ for an arbitrary binary relation and $\Box$ ($\Dia$) for its corresponding normal universal (existential) modality, \autoref{tb:property.axiom} lists some of these relational properties (also called \emph{frame conditions}), together with the formulas that characterise them  (and thus the modal property they imply) as well as the formulas' intuitive epistemic/doxastic reading.\footnote{More precisely, a \emph{frame} (a model without the valuation) has the given relational property if and only if the given formula is valid in the frame (i.e., it is true in any world of the frame under any valuation).}

\begin{table}[t]
  \begin{footnotesizectabular}{l@{\quad}l}
    \toprule
    \textbf{Frame condition}                                       & \textbf{Characterising formula} \\
    \midrule
    \emph{seriality} ($l$):                                        & \emph{consistency}: \\
    $\forall w \in W \; \exists u \in W \bdot wSu$                 & $\Box \varphi \rightarrow \Dia \varphi$ \\
    \midrule
    \emph{reflexivity} ($r$):                                      & \emph{truthfulness of knowledge/belief}: \\
    $\forall w \in W \bdot wSw$                                    & $\Box \varphi \rightarrow \varphi$ \\
    \midrule
    \emph{transitivity} ($t$):                                     & \emph{positive introspection}: \\
    $\forall w,u,v \in W \bdot ((wSu \;\&\; uSv) \Rightarrow wSv)$ & $\Box \varphi \rightarrow \Box \Box \varphi$ \\
    \midrule
    \emph{symmetry} ($s$):                                         & \emph{truthfulness of possible knowledge/belief}: \\
    $\forall w,u \in W \bdot (wSu \Rightarrow uSw)$                & $\Dia \Box \varphi \rightarrow \varphi$ \\
    \midrule
    \emph{Euclidicity} ($e$):                                      & \emph{negative introspection}: \\
    $\forall w,u,v \in W \bdot ((wSu \;\&\; wSv) \Rightarrow uSv)$ & $\neg \Box \varphi \rightarrow \Box \neg \Box \varphi$ \\
    \bottomrule
  \end{footnotesizectabular}
  \caption{Relational properties and their characterising modal formula.}
  \label{tb:property.axiom}
\end{table}

\smallskip

Now, Definitions \ref{def:DA}, \ref{def:consistency} and \ref{def:model} show that the semantic object through which $\dcaut{G}$ is semantically interpreted (the accessibility relation $\crel{G}$) is given in terms of the semantic objects through which $\beli{a}{}$ (for $a \in G$) is semantically interpreted (the relations $R_a$ for $a \in G$). In other words, semantically, the group epistemic/doxastic notion is defined in terms of the epistemic/doxastic notion for the members of the group. In cases like this, a natural and interesting question is to which extent a given property of individual knowledge/belief `carries over' to the notion of group knowledge/belief under discussion.

When both the individual notion and the group notion have relational semantics,\footnote{For positive and negative examples, while distributed and common knowledge/belief have relational semantics, the \emph{somebody knows} of \citet{agotnes2021-somebody} does not.} the question becomes whether the relevant relational property is \emph{preserved} under the operations that define the accessibility relation for the group. For example, it is well-known that if all the individual relations of the members of a group are equivalence (reflexive, transitive and Euclidean) relations, then so are the relations for the group's (standard) distributed \autorefp{def:model} and common knowledge/belief (the transitive closure of the union of the individual relations for the group's members). However, the relation for general knowledge/belief (the union of the individual relations of the group's members) might be neither transitive nor Euclidean, even when the relations of all the groups' members are.\footnote{A group has general knowledge/belief of $\varphi$ if and only if everybody in the group knows/believes $\varphi$.} A systematic study of the preservation of (all possible combinations of) the properties in \autoref{tb:property.axiom} is given, for different epistemic/doxastic notions for groups, in \citet{agotnes2021}. A general conclusion is that combinations of properties are typically \emph{not} preserved, even for standard group notions.

\medskip

Since cautious distributed belief has relational semantics, one can follow this idea and investigate systematically whether properties on the relations $R_a$ for all agents $a \in G$ are preserved for the relation on which cautious distributed belief is defined ($\crel{G}$; \autoref{def:consistency}). First, here it is the formal definition of the notion of preservation.

\begin{defi}[Preservation] \label{def:preserves.caut}
  Given a belief model $\M$ and a combination of frame conditions $\msf$ (i.e., $\msf \subseteq \set{l,r,t,s,e}$), we say that \emph{$\msf$ is preserved for cautious distributed belief in $\M$} (alternatively, that \emph{cautious distributed belief preserves $\msf$ in $\M$}) when, for any group $G \subseteq A$, the relation $\crel{G}$ \autorefp{def:consistency} satisfies $\msf$ whenever $R_a$ satisfies $\msf$ for every $a\in G$. A combination of frame conditions is preserved for cautious distributed belief on a \emph{class of models} if and only if it is preserved in every model in that class.
\end{defi}

As a first observation note that, for singleton groups, cautious distributed belief preserves all properties: if $G$ is a singleton $\set{a}$, then the cautious distributed belief relation $\crel{\set{a}}$ is identical to $a$'s individual relation $R_a$. For groups of at least two agents we have the following.

\begin{prop}\label{pro:DA:rel-properties}
  Take $G \subseteq A$ with $\card{G}\geqslant 2$. Then, cautious distributed belief
  \begin{enumerate}
    \item preserves \emph{seriality} on the class of all models;

    \item preserves \emph{reflexivity} on the class of all models; \label{pro:DA:rel-properties:2}

    \item
    \begin{enumerate}[leftmargin=1.5em]
      \item does not preserve \emph{transitivity} on the class of $\cond$ models, for $\cond \subseteq \set{l,e}$;
      \item preserves \emph{transitivity} on the class of reflexive models;
      \item preserves \emph{transitivity} on the class of symmetric models;
    \end{enumerate}

    \item
    \begin{enumerate}[leftmargin=1.5em]
      \item does not preserve \emph{symmetry} on the class of $\cond$ models, for $\cond \subseteq \set{t,e}$;
      \item does not preserve \emph{symmetry} on the class of $\cond$ models, for $\cond \subseteq \set{l}$;
      \item preserves \emph{symmetry} on the class of serial and
        Euclidian models;
      \item preserves \emph{symmetry} on the class of reflexive models;
    \end{enumerate}

    \item
    \begin{enumerate}[leftmargin=1.5em]
      \item preserves \emph{Euclidicity} on the class of serial and symmetric models;
      \item does not preserve \emph{Euclidicity} on the class of $\cond$ models, for $\cond \subseteq \set{l,t}$;
      \item does not preserve \emph{Euclidicity} on the class of $\cond$ models, for $\cond \subseteq \set{t,s}$;
      \item preserves \emph{Euclidicity} on the class of reflexive models.
    \end{enumerate}
  \end{enumerate}
  \begin{proof}
    See the \hyperref[pro:DA:rel-properties:proof]{appendix}.  
  \end{proof}
\end{prop}

The previous proposition indicates, for any combination of the five discussed frame conditions, exactly which properties are preserved in each case. While there are 32 such combinations, many of them are equivalent. \autoref{tab:frameprops} in the appendix shows an overview of equivalent combinations as well as generic names for the corresponding modal logics.

\smallskip

Summarising, \autoref{pro:DA:rel-properties} shows that seriality and reflexivity are each preserved without additional assumptions. Then, while symmetry and Euclidicity are both preserved in the presence of reflexivity, transitivity is preserved in the presence of either reflexivity or symmetry. Thus, just as with \emph{individual} belief, cautious distributed belief is truthful in reflexive models, and it is consistent in serial models. However, it does not need to be introspective (neither positively nor negatively), even when the model has the frame condition (transitivity and Euclidicity, respectively). As belief is typically assumed to be a KD45 modality (see \autoref{tab:frameprops} for the corresponding frame conditions), this tells us that cautious distributed belief is not `proper' belief, even when the individual beliefs of the group's members are (similar to standard distributed knowledge/belief; \citealp{agotnes2021}). The classes of models on which cautious distributed knowledge/belief is `well behaved' (i.e., \emph{all} properties of belief are preserved) is (K), D, T, B, S4 and S5.\footnote{The class D is included, so cautious distributed belief is consistent when all the members of the group are consistent.} For comparison, while standard distributed knowledge/belief is `well behaved' on (K), T, B, S4 and S5; common knowledge/belief is `well-behaved' on (K), S4 and S5. This means that (K), S4 and S5 are the only classes of models in which all these three notions of group knowledge/belief are well behaved. However, cautious and standard distributed belief coincide on reflexive models (\autoref{pro:DA:defDG} below), so, excluding K, there is no class of models in which the three notions are `well-behaved' while remaining different.

\subsection{Relationship between \texorpdfstring{$\dcaut{}$}{DA} and \texorpdfstring{$\dstan{}$}{D} and between \texorpdfstring{$\lang{\dcaut{}{}}$}{DA} and \texorpdfstring{$\lang{\dstan{}{}}$}{D}}\label{sub:DAvsD}

The proposition below relates $\dstan{}$ and $\dcaut{}$. First, $\dcaut{}$ is definable in terms of $\dstan{}$ and Boolean operators. Second, cautious distributed belief implies distributed belief, but not the other way around. In a picture,
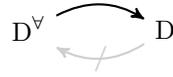
\begin{center}
  \begin{tikzpicture}[node distance = 2em and 3em, ->]
    \node[modality] (DA) {\dcaut{}{}};
    \node[modality, right = of DA] (D) {\dstan{}{}};

    \path (DA) edge [bend left = 30pt] (D)
          (D)  edge [bend left = 30pt, verylightgray] node [sloped] {/} (DA);
  \end{tikzpicture}
\end{center}
Third, both notions coincide when the accessibility relations are reflexive.

\begin{prop} \label{pro:DA:defDG}
  ~
  \begin{enumerate}
    \item $\displaystyle \vDash \dcaut{G}{\varphi} \leftrightarrow \bigwedge_{\emptyset \neq G' \subseteq G} \Big( \big( \lnot D_{G'} \bot \land \bigwedge_{G' \subset H \subseteq G} D_{H} \bot \big) \rightarrow D_{G'}\varphi \Big)$.\label{pro:DA:defDG:1}

    \item $\vDash \dcaut{G}{\varphi} \limp \dstan{G}{\varphi}$ \;but\; $\not\vDash \dstan{G}{\varphi} \limp \dcaut{G}{\varphi}$.\label{pro:DA:defDG:2}

    \item Let \class{R} be the class of reflexive belief models. Then, $\class{R} \vDash \dcaut{G}{\varphi} \leftrightarrow D_G \varphi$.\label{pro:DA:defDG:3}
  \end{enumerate}
  \begin{proof}
    Take any model $\M = \langle W,R,\val \rangle$, any $w \in W$ and any group $G \subseteq A$.
    \begin{enumerate}
      \item Suppose $\mps{w} \dcaut{G}{\varphi}$. By definition, this is the case if and only if every $\maxatin{G'}{w}{G}$ is such that $\mps{w} D_{G'} \varphi$. But $\maxatin{G'}{w}{G}$ (i.e., $G'$ is a maximally consistent subgroup of $G$ at $w$) is equivalently stated as $\mps{w} \lnot D_{G'} \bot \land \bigwedge_{G' \subset H \subseteq G} D_{H} \bot$.\footnote{Note: this relies on the fact that $G$ is finite (because $A$ is finite).} Then, the previous is the case if and only if $\mps{w} \bigwedge_{G' \subseteq G} \big( ( \lnot D_{G'} \bot \land \bigwedge_{G' \subset H \subseteq G} D_{H} \bot ) \rightarrow D_{G'}\varphi \big)$.

      \item The first follows from the fact that $\gcs{G}{w} \subseteq \cgcs{G}{w}$ for every group $G$ in every world $w$, which is proved by cases. If $G$ is consistent at $w$, then it is its own single maximally consistent set, so $\gcs{G}{w} = \cgcs{G}{w}$. Otherwise, $\gcs{G}{w}$ is $\emptyset$ and thus a subset of any set. For the second, \autoref{exa:DA:1} is a pointed model in which $\dstan{G_1}{\bot}$ and yet $\lnot \dcaut{G_1}{\bot}$.

      \item Immediate, as $\gcs{G}{w} = \cgcs{G}{w}$ holds for any $\M$ in $\class{R}$, any $w \in \dom(\M)$ and any group $G \subseteq A$ (see the \hyperref[pro:DA:rel-properties:proof]{proof} of \autoref{pro:DA:rel-properties}.\ref{pro:DA:rel-properties:2}).
    \end{enumerate}
  \end{proof}
\end{prop}


Having shown the relationship between $\dstan{}$ and $\dcaut{}$ (\autoref{pro:DA:defDG:2} just above), it is time to compare their respective languages $\lang{\dstan{}{}}$ and $\lang{\dcaut{}{}}$. Using \autoref{pro:DA:defDG:1} in \autoref{pro:DA:defDG}, one can define a translation that takes any formula in $\lang{\dcaut{}{}}$ and returns a semantically equivalent formula in $\lang{\dstan{}{}}$. This already establishes a connection between the languages under discussion.


\begin{coro}
  $\lang{\dstan{}{}}$ is at least as expressive as $\lang{\dcaut{}{}}$ (in symbols: $\lang{\dcaut{}{}} \preccurlyeq \lang{\dstan{}{}}$).
\end{coro}

A question remains: does $\lang{\dstan{}{}} \preccurlyeq \lang{\dcaut{}{}}$ hold ($\lang{\dcaut{}{}}$ is at least as expressive as $\lang{\dstan{}{}}$) and thus $\lang{\dcaut{}{}} \approx \lang{\dstan{}{}}$ (the languages are equally expressive), or does $\lang{\dstan{}{}} \not\preccurlyeq \lang{\dcaut{}{}}$ hold ($\lang{\dcaut{}{}}$ is \emph{not} at least as expressive as $\lang{\dstan{}{}}$) and thus $\lang{\dcaut{}{}} \prec \lang{\dstan{}{}}$ ($\lang{\dstan{}{}}$ is \emph{strictly} more expressive than $\lang{\dcaut{}{}}$)?

\medskip

As mentioned (immediately below \autoref{def:expressivity}), one can prove that $\lang{1} \not\preccurlyeq \lang{2}$ holds by providing two pointed models that satisfy exactly the same formulas in $\lang{2}$, and yet can be distinguished by a formula in $\lang{1}$. For arguing that two pointed models cannot be distinguished by formulas of a given language, one normally uses a semantic form of similarity between models. For normal modal logics with relational semantics, the notion of \emph{bisimulation} \citep[Definition 2.18]{Blackburn2002} plays this role: it is well-known that, if $\Box$ is a normal modality, formulas of the form $\Box\varphi$ are invariant under a notion of bisimulation that works with the binary relation on which $\Box$ is semantically interpreted \citep[Theorem 2.20]{Blackburn2002}.\footnote{This assumes that only normal modalities are involved in the language.} Thus, for example, formulas of the basic multi-agent epistemic language are invariant under a form of bisimulation that works with the accessibility relation of each one of the agents. Moreover, for a language that extends the previous one with the modality for standard distributed belief (the language $\lang{\dstan{}{}}$), one simply requires a bisimulation that works on the relations that are now relevant: the relations $R_G$ for each group $G$ \autorefp{def:model}. In \citet{roelofsen2007}, this is called \emph{collective bisimulation}.

\smallskip

Since $\dcaut{}$ has relational semantics, one can define an adequate form of bisimulation by working with the relations $\crel{G}$ instead. The following definition, for the language $\lang{\dcaut{}{}}$, is a standard bisimulation that uses $\crel{G}$, with the latter expanded according to its definition \autorefp{def:consistency}.

\begin{defi}[$\lang{\dcaut{}{}}$-Bisimulation] \label{def:DA:bisim}
  Let $\M=\langle W, R, \val \rangle$ and $\M'=\langle W',R',\val' \rangle$ be two belief models. A non-empty relation $Z \subseteq W \times W'$ is a \emph{$\lang{\dcaut{}{}}$-bisimulation} between $\M$ and $\M'$ if and only if $Zww'$ implies all of the following.
  \begin{bisimitemize}
    \item \textbf{Atom.} For all $p \in P$: $w \in \val(p)$ if and only if $w' \in \val'(p)$.

    \item \textbf{Forth.} For all groups $G \subseteq A$ and all $u \in W$: if there is $\maxatin{H}{w}{G}$ such that $u \in \gcs{H}{w}$, then there are $H' \subseteq G$ and $u' \in W'$ such that $\maxatin{H'}{w'}{G}$, $u'\in \gcs{H'}{w'}$ and $Zuu'$.

    \item \textbf{Back.} For all groups $G \subseteq A$ and all $u' \in W'$: if there is $\maxatin{H'}{w'}{G}$ such that $u' \in \gcs{H'}{w'}$, then there are $H \subseteq G$ and $u \in W$ such that $\maxatin{H}{w}{G}$, $u \in \gcs{H}{w}$ and $Zuu'$.
  \end{bisimitemize}
  Write $\M,w \bidcautover{Z} \M',w'$ when $Z$ is a $\lang{\dcaut{}{}}$-bisimulation between $\M$ and $\M'$ with $Zww'$. Write $\M,w \bidcaut \M',w$ when there is such a bisimulation $Z$.
\end{defi}

It is worth reflecting on this particular instance of the notion of bisimulation. As usual, $\lang{\dcaut{}{}}$-bisimilar worlds should satisfy the same atoms. Then, if one of them has a `relevant successor' $u$, the other should also have a `relevant successor' $u'$ and, moreover, these successors should be $\lang{\dcaut{}{}}$-bisimilar. The only difference between a $\lang{\dcaut{}{}}$-bisimulation and bisimulation for other logics is what `a relevant successor' means. In a multi-agent standard bisimulation, a `relevant successor' is any world that can be reached through the relation $R_a$ of some agent $a \in A$. In a collective bisimulation, a `relevant successor' is any world that can be reached through the relation $R_G$ of some group $G \subseteq A$. In the just defined $\lang{\dcaut{}{}}$ bisimulation, a `relevant successor' is any world that belongs to the conjecture set of some maximally consistent subgroup of $G$, for some (recall: non-empty) group $G \subseteq A$.\footnote{Note: a group that is inconsistent in a given world does not need to be inconsistent in $\lang{\dcaut{}{}}$-bisimilar worlds (see the models in the proof of \autoref{fct:DA.less.D} below). This is different from the collective bisimulation case, under which a group that is inconsistent in a given world must be also inconsistent at any collectively bisimilar one.} This guarantees that every world in $W$ that is relevant for \caut distributed belief in $(\M,w)$ has a `matching' world in $W'$ that is relevant for \caut distributed belief in $(\M',w')$ (and vice versa). An example of $\lang{\dcaut{}{}}$-bisimilar models is given in the proof of \autoref{fct:DA.less.D} below.

\smallskip

One can now work towards answering the question posed above. First, the formal definition of equivalence w.r.t. $\lang{\dcaut{}{}}$.

\begin{defi}[$\lang{\dcaut{}{}}$-equivalence]
  Two pointed models $\M,w$ and $\M',w'$ are $\lang{\dcaut{}{}}$-equivalent (notation: $\M,w \eqdcaut \M',w'$) if and only if, for every $\varphi \in \lang{\dcaut{}{}}$,
  \begin{ctabular}{r@{\quad{if and only if}\quad}l}
    $\M, w \vDash \varphi$ & $\M',w' \vDash \varphi$.
  \end{ctabular}
  When the models are clear from context, we will write simply $w \eqdcaut w'$.
\end{defi}

The following two results are direct consequences of well-known results of the notion of bisimulation (Theorem 2.20 and Theorem 2.24 in \citealp{Blackburn2002}, respectively).

\begin{prop}[$\lang{\dcaut{}{}}$-Bisimilarity implies $\lang{\dcaut{}{}}$-equivalence]\label{pro:DA:invar}
  Let $\M,w$ and $\M',w'$ be pointed belief models. Then,
  \begin{ctabular}{r@{\qquad{implies}\qquad}l}
    $\M,w \bidcaut \M',w'$ & $\M,w \eqdcaut \M',w'$.
  \end{ctabular}
\end{prop}

\begin{prop}[$\lang{\dcaut{}{}}$-Equivalence implies $\lang{\dcaut{}{}}$-bisimilarity]\label{pro:DA:implies}
  Let $\M,w$ and $\M',w'$ be \emph{individually image-finite} pointed belief models.\footnote{\label{fnt:in-img-fin}A belief model $\M=\langle W,R,v \rangle$ is \emph{individually image-finite} iff $\gcs{a}{w}$ is finite for every agent $a \in A$ and every $w \in \dom(\M)$. Note how this is equivalent to $\gcs{G}{w}$ and $\cgcs{G}{w}$ being finite for every $G \subseteq A$ and every $w \in \dom(\M)$.} 
  Then,
  \begin{ctabular}{r@{\qquad{implies}\qquad}l}
    $\M,w \eqdcaut \M',w'$ & $\M,w \bidcaut \M',w'$.
  \end{ctabular}
\end{prop}

We can now answer the question above.

\begin{fact}\label{fct:DA.less.D}
  $\lang{\dcaut{}{}}$ is \emph{not} at least as expressive as $\lang{\dstan{}{}}$ (in symbols: $\lang{\dstan{}{}} \not\preccurlyeq \lang{\dcaut{}{}}$).
  \begin{proof}
    Consider the belief models shown below.
    
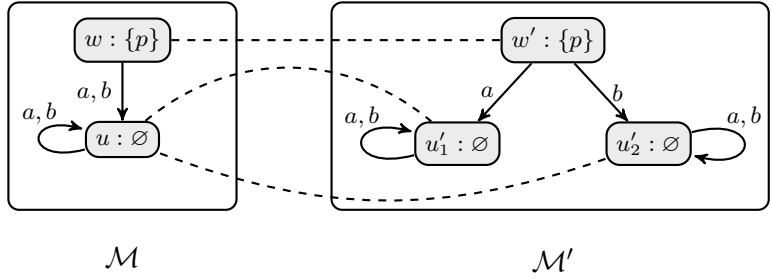
\begin{center}
      \begin{tikzpicture}[node distance = 2em and 0em, ->]
        \node  at (0,0) [world] (w) {w:\set{p}};
        \node[world, below = of w] (u) {u:\emptyset};
        \path (w) edge node [left, label edge] {$a,b$} (u);
        \path (u) edge [loop left] node [above = 2pt, label edge] {$a,b$} ();
        \node [below = 6em of w] {$\M$};
        \draw [frame rectangle] (-1.5,-2.25) rectangle (1.5, 0.5);

        \node[world, right = 11.25em of w] (w') {w':\set{p}};
        \node[world, below left = of w'] (u'1) {u'_1:\emptyset};
        \node[world, below right = of w'] (u'2) {u'_2:\emptyset};
        \path (w') edge node [left, label edge] {$a$} (u'1)
                   edge node [right, label edge] {$b$} (u'2)
              (u'1) edge [loop left] node [above = 2pt, label edge] {$a,b$} ()
              (u'2) edge [loop right] node [above = 2pt, label edge] {$a,b$} ();
        \node [below = 6em of w'] {$\M'$};

        \path (w) edge [bisim] (w')
              (u) edge [bisim, bend left = 40pt] (u'1)
                  edge [bisim, bend right = 20pt] (u'2);
        \draw [frame rectangle] (2.75,-2.25) rectangle (8.5, 0.5);
      \end{tikzpicture}
    \end{center}
    Use $\maxat{G}{v}$ to denote all subgroups of $G$ that are maximally consistent at a given world $v$. The dashed edges define a $\lang{\dcaut{}{}}$-bisimulation between $\M$ and $\M'$, as the pairs $(w,w')$, $(u,u'_1)$ and $(u,u'_2)$ satisfy the three clauses.
    \begin{compactitemize}
      \item $\bm{(w,w')}$. The \textbf{atom} clause is immediate. Now \textbf{forth}. For $G = \set{a}$, note that $\maxat{\set{a}}{w} = \set{\set{a}}$ and thus $\cgcs{\set{a}}{w} = \set{u}$. But then $\maxat{\set{a}}{w'} = \set{\set{a}}$ and thus $\cgcs{\set{a}}{w'} = \set{u'_1}$; moreover, $Zuu'_1$. The case for $G = \set{b}$ is analogous. For $G = \set{a,b}$, note that $\maxat{\set{a,b}}{w} = \set{\set{a,b}}$ and thus $\cgcs{\set{a,b}}{w} = \set{u}$. But then $\maxat{\set{a,b}}{w'} = \set{\set{a},\set{b}}$ and thus $\cgcs{\set{a,b}}{w'} = \set{u'_1, u'_2}$; moreover, $Zuu'_1$ and $Zuu'_2$. The \textbf{back} clause follows a similar pattern.

      \item $\bm{(u,u'_1)}$. The \textbf{atom} clause is immediate. Consider \textbf{forth}. For $G = \set{a}$, note that $\maxat{\set{a}}{u} = \set{\set{a}}$ and thus $\cgcs{\set{a}}{u} = \set{u}$. But then $\maxat{\set{a}}{u'_1} = \set{\set{a}}$ and thus $\cgcs{\set{a}}{u'} = \set{u'_1}$; moreover, $Zuu'_1$. The case for $G = \set{b}$ is analogous. For $G = \set{a,b}$, note that $\maxat{\set{a,b}}{u} = \set{\set{a,b}}$ and thus $\cgcs{\set{a,b}}{u}= \set{u}$. But then $\maxat{\set{a,b}}{u'_1} = \set{\set{a,b}}$ and thus $\cgcs{\set{a,b}}{u'_1} = \set{u'_1}$; moreover, $Zuu'_1$. The \textbf{back} clause follows a similar pattern.

      \item $\bm{(u,u'_2)}$.  As the previous case.
    \end{compactitemize}
    Thus, $M,w \bidcaut \M',w'$ and hence $\M,w \eqdcaut \M',w'$. However, the pointed models can be distinguished by a formula in $\lang{\dstan{}{}}$, as $\mpf{w} \dstan{\set{a,b}}{\bot}$ and yet $\M',w' \vDash \dstan{\set{a,b}}{\bot}$. Therefore $\lang{\dstan{}{}} \not\preccurlyeq \lang{\dcaut{}{}}$.
  \end{proof}
\end{fact}

Here are two observations about the pointed models that prove \autoref{fct:DA.less.D}.
\begin{itemize}
  \item The belief models are serial, transitive and Euclidean: the kind of models one normally uses for representing beliefs.

  \item The fact that the pointed models can be distinguished by $\lang{\dstan{}{}}$ can be backed up semantically: no \emph{collective} bisimulation between these models contains the pair $(w,w')$ because $\gcs{\set{a,b}}{w} \neq \emptyset$ (i.e., some world is $R_{\set{a,b}}$-reachable from $w$) and yet $\gcs{\set{a,b}}{w'} = \emptyset$ (i.e., no world is $R'_{\set{a,b}}$-reachable from $w'$).
\end{itemize}

With this, we can reach our final goal.

\begin{coro}\label{coro:DA.less.D}
  $\lang{\dstan{}{}}$ is strictly more expressive than $\lang{\dcaut{}{}}$ (symbols: $\lang{\dcaut{}{}} \prec \lang{\dstan{}{}}$).
\end{coro}

Thus, $\lang{\dcaut{}{}}$ can `see' strictly less than what $\lang{\dstan{}{}}$ can. The proposition below shows that the group inconsistency constant $\inc{G}$ introduced before is what the former needs to `see' exactly as much as the latter.

\begin{prop}\label{prp:equal}
  $\lang{\dcaut{}{}, \inc{}}$ and $\lang{\dstan{}{}}$ are equally expressive (symbols: $\lang{\dcaut{}{}, \inc{}} \approx \lang{\dstan{}{}}$).
  \begin{proof}
    For proving $\lang{\dcaut{}{}, \inc{}} \preccurlyeq \lang{\dstan{}{}}$, note that $\vDash \inc{G} \ldimp \dstan{G}{\bot}$. Thus, both $\inc{G}$ and $\dcaut{G}{}$ are definable in $\lang{D}$ (for the latter, recall \autoref{pro:DA:defDG}).

    For proving $\lang{\dstan{}{}} \preccurlyeq \lang{\dcaut{}{}, \inc{}}$, note that $\vDash \dstan{G}{\varphi} \ldimp (\inc{G} \lor \dcaut{G}{\varphi})$. {\prooflr} Suppose $\mps{w} \dstan{G}{\varphi}$, so $\mps{u}\varphi$ for every $u \in \gcs{G}{w}$. Assume further that $\mpf{w} \inc{G}$. Then, $\maxat{G}{w} = \set{G}$ and thus $\mps{w}\dcaut{G}{\varphi}$. {\proofrl} Proceed by contraposition: suppose $\mpf{w} \dstan{G}{\varphi}$. Then, there is $u \in \gcs{G}{w}$ such that $\mpf{u} \varphi$. Thus, $\gcs{G}{w} \neq \emptyset$, so $\maxat{G}{w} = \set{G}$. From the former, $\mpf{w} \inc{G}$; from $u \in \gcs{G}{w}$ and the latter, $\mpf{w} \dcaut{G}{\varphi}$. Thus, $\mpf{w} \inc{G} \lor \dcaut{G}{\varphi}$. From this validity, $\dstan{G}{}$ is definable in $\lang{\dcaut{}{}, \inc{}}$.
  \end{proof}
\end{prop}

\section{Bold distributed belief}\label{sec:DE}


\subsection{Semantic interpretation}
As mentioned in the introduction, the idea behind cautious distributed belief is that, although a group as a whole might be inconsistent, there might be consistent subgroups. Among them, the maximally consistent ones are particularly relevant. \emph{Cautious} distributed belief uses these subgroups in a `safe' way: $\dcaut{G}{\varphi}$ holds if and only if \emph{all} of $G$'s maximally consistent subgroups believe $\varphi$ distributively. The notion of \emph{bold} distributed belief discussed in this section (modality: $\dbold{}$) follows a `more adventurous' idea: $\dbold{G}{\varphi}$ holds if and only if \emph{at least one} of $G$'s maximally consistent subgroups believe $\varphi$ distributively. The difference can be seen as different assumptions about how groups resolve disagreements.

\begin{defi}[Bold distributed belief]\label{def:DE}
  Let $(\M, w)$ be a pointed belief model with $\M = \langle W,R,\val \rangle$; take group $G \subseteq A$. Then,
  \begin{center}
    \begin{tabular}[b]{l@{\qquad{iff}\qquad}l}
      $\M,w \vDash \dbold{G}{\varphi}$ & $\exists \maxatin{G'}{w}{G}$: $\M,w \vDash \dstan{G'}{\varphi}$ \\
    \end{tabular}
  \end{center}
  Validity (in a class of belief models) is defined and denoted as before.
\end{defi}

It is useful to expand the definition of distributed belief in the previous definition. By doing so, the semantic clause for $\dbold{G}{}$ becomes
\begin{center}
  \begin{tabular}{l@{\qquad{iff}\qquad}l}
    $\M,w \vDash \dbold{G}{\varphi}$ & $\exists \maxatin{G'}{w}{G}$, $\forall w' \in \gcs{G'}{w}$: $\M,w' \vDash \varphi$, \\
  \end{tabular}
\end{center}
thus making explicit the quantification pattern: first an existential one over maximally consistent subgroups, and then a universal one over the worlds in this group's conjecture set.

\begin{exam1}\label{exa:DE:1}
  Consider again the model from \autoref{exa:DA:1} in \autoref{sec:DA}: 
    \begin{center}
    \begin{tikzpicture}[frame rectangle]
      \node[world, double] (w) {w_1:\set{p,q}};
      \node[world, below left = 3em and 6em of w] (u1) {w_2:\set{p}};
      \node[world, below = 3em of w] (u2) {w_3:\set{q}};
      \node[world, below right = 3em and 6em of w] (u3) {w_4:\set{p}};

      \path (w)  edge node [above, label edge] {$a$} (u1)
                 edge node [left, label edge] {$b$} (u2)
                 edge [loop right] node [right, label edge] {$b,c$} ()
                 edge node[above, label edge] {$c$} (u3)
            (u1) edge [loop left] node [above = 5pt, label edge] {$a,b,c$} ()
            (u2) edge [loop right] node [above = 5pt, label edge] {$a,b,c$} ()
            (u3) edge [loop right] node [above = 5pt, label edge] {$a,b,c$} ();
    \end{tikzpicture}
  \end{center}
  Recalling that, for $G_1=\set{a,b}$, the conjecture sets of the maximally consistent subgroups are $\set{\set{w_2},\set{w_1,w_3}}$, we see that the group has bold distributed beliefs that are not cautious distributed beliefs: $\mps{w_1} \dbold{G_1}{p} \land \dbold{G_1}{q}$. Now, since $\extm{q} \cap \set{w_1}=\emptyset$, it is also the case that $\mps{w_1} \dbold{G_1}{\lnot q}$, the group has bold distributed belief both in $q$ and its negation. However, importantly, the group does not have bold distributed belief in the absurd, nor in the conjunction of $q$ and its negation: $\M,w_1 \nvDash \dbold{G_1}{\bot}$ and $\M,w_1 \nvDash \dbold{G_1}{(q \land \lnot q)}$. If we look at the group $G_2=\set{a,b,c}$, the situation is similar. Like before however, $b$ and $c$ are consistent, and their information can be refined, so the relevant conjecture sets are $\set{\set{w_2},\set{w_1}}$, and so we also get $\mps{w_1}\vdash \dbold{G_2}{(p \land q)}$.
\end{exam1}

Now, here is the most important observation about $\dbold{}{}$: different from $\dcaut{}{}$, it does not have relational semantics.

\begin{fact}\label{fct:DE:no.rel}
  The modality $\dbold{}{}$ does not have relational semantics.
  \begin{proof}
    A normal unary modality with relational semantics is either a standard `box' or a standard `diamond': its semantic interpretation quantifies (universally or existentially, respectively) over the set of worlds that can be reached by the given binary relation. But consider the two following observations. First, $\dbold{}{}$ does \emph{not} distribute over implications: $\dbold{G}{(\varphi \limp \psi)} \limp (\dbold{G}{\varphi} \limp \dbold{G}{\psi})$ is \emph{not} valid. For this, consider the model $\M_1$ below on the left. For the group $\set{a,b}$, the maximally consistent subsets at $w_1$ are $\set{a}$ and $\set{b}$. Since $\M_1, w_1 \vDash \dstan{a}{(p \limp q)} \land \dstan{b}{p}$, one can see that $\M_1, w_1 \vDash \dbold{\set{a,b}}{(p \limp q)} \land \dbold{\set{a,b}}{p}$. Nevertheless, $\M_1, w_1 \vDash \lnot \dstan{a}{q} \land \lnot \dstan{b}{q}$, and therefore $\M_1, w_1 \not\vDash \dbold{\set{a,b}}{q}$.
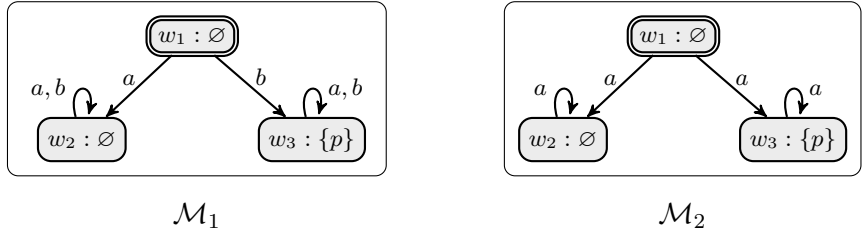
    \begin{ctabular}{c@{\qquad\qquad}c}
      \begin{tikzpicture}[frame rectangle]
        \node[world, double] (w1) {w_1:\emptyset};
        \node[world, below left  = 2em and 0em of w] (w2) {w_2:\emptyset\vphantom{\set{p,q}}};
        \node[world, below right = 2em and 0em of w] (w3) {w_3:\set{p}};

        \path (w1) edge node [above, label edge, pos=0.65] {$a$} (w2)
                   edge node [above, label edge, pos=0.65] {$b$} (w3)
              (w2) edge [loop above] node [left=2pt, label edge] {$a,b$} ()
              (w3) edge [loop above] node [right=2pt, label edge] {$a,b$} ();
      \end{tikzpicture}
      &
      \begin{tikzpicture}[frame rectangle]
        \node[world, double] (w1) {w_1:\emptyset};
        \node[world, below left  = 2em and 0em of w] (w2) {w_2:\emptyset\vphantom{\set{p,q}}};
        \node[world, below right = 2em and 0em of w] (w3) {w_3:\set{p}};

        \path (w1) edge node [above, label edge, pos=0.65] {$a$} (w2)
                   edge node [above, label edge, pos=0.65] {$a$} (w3)
              (w2) edge [loop above] node [left=2pt, label edge] {$a$} ()
              (w3) edge [loop above] node [right=2pt, label edge] {$a$} ();
      \end{tikzpicture}
      \\
      $\M_1$ & $\M_2$
    \end{ctabular}
    For the second observation, define the dual of $\dbold{}{}$ in the standard way, $\ddbold{G}{\varphi} := \lnot \dbold{G}{\lnot \varphi}$, so
    \begin{center}
      \begin{tabular}[b]{l@{\qquad{iff}\qquad}l}
        $\M,w \vDash \ddbold{G}{\varphi}$ & $\M,w \vDash \lnot \dbold{G}{\lnot \varphi}$ \\
                                          & not $\exists \maxatin{G'}{w}{G}$: $\M,s \vDash \dstan{G'}{\lnot\varphi}$ \\
                                          & $\forall \maxatin{G'}{w}{G}$: $\M,w \not\vDash \dstan{G'}{\lnot\varphi}$ \\
                                          & $\forall \maxatin{G'}{w}{G}$: $\M,w \vDash \ddstan{G'}{\varphi}$ \\
                                          & $\forall \maxatin{G'}{w}{G}$, $\exists w' \in \gcs{G'}{w}$: $\M,w' \vDash \varphi$. \\
      \end{tabular}
    \end{center}
    Now, $\ddbold{}{}$ does not distribute over implications either: $\ddbold{G}{(\varphi \limp \psi)} \limp (\ddbold{G}{\varphi} \limp \ddbold{G}{\psi})$ is not valid. For this, consider the model $\M_2$ above on the right. For the group $\set{a}$, the only maximally consistent subset at $w_1$ is $\set{a}$ itself. Since $\M_2, w_1 \vDash \ddstan{a}{(p \limp q)} \land \ddstan{a}{p}$, one can see that $\M_2, w_1 \vDash \ddbold{\set{a}}{(p \limp q)} \land \ddbold{\set{a}}{p}$. Nevertheless, $\M_2, w_1 \vDash \lnot \ddstan{a}{q}$, so $\M_2, w_1 \not\vDash \ddbold{\set{a}}{q}$.

    \smallskip

    The first observation tells us that the modality $\dbold{}{}$ is not a universal (`box') normal modality. The second observation tells us that neither is $\ddbold{}{}$, and thus its dual $\dbold{}{}$ is not an existential (`diamond') normal modality. Thus, $\dbold{}{}$ does not have relational semantics.
  \end{proof}
\end{fact}

In order to show that $\dbold{}{}$ does not have relational semantics, the previous proposition showed that neither $\dbold{}{}$ nor its modal dual $\ddbold{}{}$ distribute over implications. The reader might then wonder about other properties a normal modality has. Here are some answers.



\begin{prop}\label{pro:DE:properties}
  ~
  \begin{enumerate}
    \item $\vDash \dbold{G}{(\varphi \land \psi)} \limp (\dbold{G}{\varphi} \land \dbold{G}{\psi})$.

    \item\label{pro:DE:properties:conj-int} $\nvDash (\dbold{G}{\varphi} \land \dbold{G}{\psi}) \rightarrow \dbold{G}{(\varphi \land \psi)}$.

    \item $\vDash \varphi \leftrightarrow \psi$ \; implies \; $\vDash \dbold{G}{\varphi} \leftrightarrow \dbold{G}{\psi}$.

    \item $\vDash \varphi\rightarrow \psi$ \; implies \; $\vDash \dbold{G}{\varphi} \rightarrow \dbold{G}{\psi}$.

    \item\label{pro:DE:properties:val} $\vDash \varphi$ \; does not imply \; $\vDash \dbold{G}{\varphi}$.
  \end{enumerate}
  \begin{proof}
    ~
    \begin{enumerate}
      \item Take an arbitrary $(\M, w)$ with $\mps{w}{\dbold{G}{(\varphi \land \psi)}}$. Then, $\exists\maxatin{G'}{w}{G}$ such that $w' \in \gcs{G'}{w}$ implies $\mps{w'}\varphi \land \psi$ and thus both $w' \in \gcs{G'}{w}$ implies both $\mps{w'}\varphi$ and $\mps{w'} \psi$. Then, $\mps{w}\dbold{G}{\varphi}$ and $\mps{w}\dbold{G}{\psi}$, and hence $\mps{w} \dbold{G}{\varphi} \land \dbold{G}{\psi}$.

      \item Consider the pointed model:
      
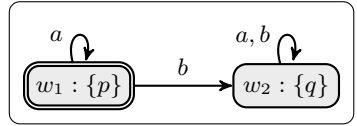
\begin{center}
        \begin{tikzpicture}[frame rectangle]
          \node[world, double] (w1) {w_1:\set{p}};
          \node[world, right  = 3em of w] (w2) {w_2:{\set{q}}};
          \path (w1) edge node [above, label edge] {$b$} (w2)
                     edge [loop above] node [left=2pt, label edge] {$a$} ()
                (w2) edge [loop above] node [left=2pt, label edge] {$a,b$} ();
        \end{tikzpicture}
      \end{center}
      The maximally consistent groups in $\set{a,b}$ at $w_1$ are $\set{a}$ and $\set{b}$. From $\gcs{\set{b}}{w_1} = \set{w_2}$ and $\mps{w_2} q$, it follows that $\mps{w_1} \dbold{\set{a,b}}{q}$; from $\gcs{\set{a}}{w_1}=\set{w_1}$ and $\mps{w_1}p$ it follows that $\mps{w_1} \dbold{\set{a,b}}{p}$. However, there is no $\maxatin{G'}{w_1}{\set{a,b}}$, such that $\forall w' \in G'$, $\mps{w'} p \land q$. Thus $\mpf{w_1} (\dbold{G}{p} \land \dbold{G}{q}) \rightarrow \dbold{G}{(p \land q)}$.

      \item Assume $\vDash \varphi \leftrightarrow \psi$. Take an arbitrary $(\M, w)$ and, from left to right, assume $\mps{w} \dbold{G}{\varphi}$. Then, $\exists \maxatin{G'}{w}{G}$ and $w' \in \gcs{G'}{w}$ implies $\mps{w'} \varphi$; thus, $w' \in \gcs{G'}{w}$ implies $\mps{w'}\psi$, so $\mps{w}\dbold{G}{\psi}$. Hence, $\mps{w} \dbold{G}{\varphi} \rightarrow \dbold{G}{\psi}$. The right-to-left direction is analogous. Since $(\M, w)$ is arbitrary, $\vDash \dbold{G}{\varphi} \leftrightarrow \dbold{G}{\psi}$

      \item The first part of the item just above.

      \item The formula $\dbold{G}{\top}$ fails exactly in those pointed models $(\M,w)$ in which $G$ has no maximally consistent subgroups, i.e., in pointed models where $\gcs{G'}{w} = \emptyset$ for all $G' \subseteq G$ (equivalently, in pointed models where all agents are inconsistent). 
    \end{enumerate}
  \end{proof}
\end{prop}

\autoref{pro:DE:properties:val} in \autoref{pro:DE:properties} states that $\dbold{G}$ does not satisfy the generalisation rule: it does not need to contain all validities. The provided argument actually characterises the situations in which that happens: when every member of the group is individually inconsistent.

\begin{coro}\label{cor:DE:tautology}
  For every group $G \subseteq A$ we have $\displaystyle \vDash \neg \dbold{G}{\top} \leftrightarrow \bigwedge_{a \in G} \beli{a}{\bot}$.
\end{coro}

Note how this characterisation mirrors the formula characterising the situations in which a group of agents has \emph{cautious} distributed belief in contradictions (\autoref{pro:DA}.\ref{pro:DA:itm:consistency}).

\medskip

Finally, even though $\dbold{}$ does not have relational semantics, the quantification pattern in its semantic clause (together with the properties it satisfies) suggests that the modality does have neighbourhood semantics \citep{Chellas1980,Pacuit2017}. This is indeed the case: for each group $G \subseteq A$, one can define a \emph{neighbourhood} function $N_G:W \to \wp(\wp(W))$ through which $\dbold{}$ can be semantically interpreted. Here we will use the `strict' neighbourhood semantics; hence, the neighbourhood of a group $G$ defined below will be \emph{monotonic}: it will contain not only the conjecture sets of every maximally consistent subgroup of $G$, but also their supersets.\footnote{Note: a similar result can be obtained by defining $N_G$ as only the conjecture sets of every maximally consistent subgroup of $G$ and then using the `loose' neighbourhood semantics ($\M,w \vDash \dbold{G}{\varphi}$ iff there is $U \in N_G(w)$ with $U \subseteq \extm{\varphi}$). See \citet{ArecesF09} for a comparison between the two semantics.}

\begin{prop}\label{pro:DE:neigh:G}
  Let $(\M, w)$ be a pointed belief model. For each group $G \subseteq A$, define the \emph{neighbourhood function $N_G:W \to \wp(\wp(W))$} as $N_G(w) := \set{ U \subseteq W \mid \gcs{G'}{w} \subseteq U \text{ for some } \maxatin{G'}{w}{G} }$. Then,
  \begin{center}
    \begin{tabular}[b]{l@{\qquad{iff}\qquad}l}
      $\M,w \vDash \dbold{G}{\varphi}$ & $\extm{\varphi} \in N_G(w)$.
    \end{tabular}
  \end{center}
  \begin{proof}
    Take an arbitrary $(\M, w)$. Then, $\mps{w}\dbold{G}{\varphi}$ iff $\exists \maxatin{G'}{w}{G}$ such that $\gcs{G'}{w} \subseteq \extm{\varphi}$ iff (by construction of $N_G$) $\extm{\varphi} \in N_G(w)$.
  \end{proof}
\end{prop}


\subsection{Some basic results about \texorpdfstring{$\dbold{}$}{DE}}\label{sub:DE:basics}

Here are three basic properties of the bold distributed belief modality. The first one states that, unlike $\dstan{}$ and $\dcaut{}$, bold distributed belief is \emph{never} inconsistent. The second indicates that, while bold distributed belief for a singleton ($\dbold{\set{a}}$) is different from individual belief ($\beli{a}$), the latter can still be expressed using $\dbold{\set{a}}$ and Boolean operators. The third one tells us that, just like $\dstan{}$ (and unlike $\dcaut{}$), the modality for bold distributed belief is group monotonic.

\begin{prop}\label{pro:DE}
  Take any groups $G \subseteq H \subseteq A$ and any $a \in A$.
  \begin{enumerate}
    \item\label{pro:DE:absurdtautology} $\vDash \lnot\dbold{G}{\bot}$.

    \item\label{pro:DE:single.agent} $\nvDash \beli{a}{\varphi} \ldimp \dbold{\set{a}}{\varphi}$ \; and yet \; $\vDash \beli{a}{\varphi} \leftrightarrow (\neg \dbold{\set{a}}{\top} \lor \dbold{\set{a}}{\varphi})$.

    \item\label{pro:DE:itm:coalition-mono} $\vDash \dbold{G}{\varphi} \rightarrow \dbold{H}{\varphi}$.
  \end{enumerate}
  \begin{proof}
    Take any $\M$ and any $w \in \dom(\M)$.
    \begin{enumerate}
      \item For $\dbold{G}{\bot}$ to hold in a given pointed model, one needs a maximally consistent subset of $G$ that has distributed belief in $\bot$. But no \emph{consistent} set of agents believes $\bot$ distributively.

      \item For the first, simply note that $\beli{a}{\bot}$ is satisfiable (when $a$ is individually inconsistent) but, as observed above, $\dbold{a}{\bot}$ is not.

      For the second, from left to right assume $\M,w \vDash \beli{a}{\varphi}$, so $\forall w' \in \gcs{a}{w}$, $\M,w'\vDash \varphi$. There are two cases. If $\gcs{a}{w}= \emptyset$, then $\M,w\vDash \neg \dbold{\set{a}}{\top}$ (from \autoref{cor:DE:tautology}). If $\gcs{a}{w}\neq \emptyset$, then, since $\maxatin{\set{a}}{w}{\set{a}}$, $\M,w \vDash \dbold{\set{a}}{\varphi}$. Thus, $M,w \vDash \neg \dbold{\set{a}}{\top} \lor \dbold{\set{a}}{\varphi}$. From right to left, assume $\M,w \nvDash \beli{a}{\varphi}$. Then, $\exists w' \in \gcs{\set{a}}{w}$ such that $\M,w' \nvDash \varphi$, which also tells us that $\set{a}$ is consistent and thus its only maximally consistent group. Then, first, $\gcs{\set{a}}{w}\neq \emptyset$ and thus $\M,w \nvDash \neg \dbold{\set{a}}{\top}$. Second, $\M,w'\nvDash \dbold{\set{a}}{\varphi}$. Thus, $M,w \nvDash \neg \dbold{\set{a}}{\top} \lor \dbold{\set{a}}{\varphi}$.

      \item Assume that $\mps{w}\dbold{G}{\varphi}$. Then $\exists \maxatin{G'}{w}{G}, \forall w' \in \gcs{G'}{w}$: $ \mps{w'}\varphi$. Take $H \supseteq G$. Then, either $\maxatin{G'}{w}{H}$ or $\exists H' \subseteq H$ such that $G' \subset H'$, $\gcs{H'}{w} \subseteq \gcs{G'}{w}$ and $\maxatin{H'}{w}{H}$. In either case $\exists \maxatin{H'}{w}{H}$ such that $\forall w' \in \gcs{H'}{w}$: $\mps{w'}\varphi$. Thus $\mps{w}\dbold{H}{\varphi}$.
    \end{enumerate}
  \end{proof}
\end{prop}

\subsection{Preserving doxastic properties}\label{sub:DE:preserving}

Just as in \autoref{sub:DA:preserving}, it is interesting to study whether individual belief properties (introspection, consistency and truthfulness) are preserved by the operation that defines bold distributed belief. In other words, if the beliefs of every agent in a group are truthful/consistent/introspective, are the bold beliefs of the group as well?

In the case of cautious distributed belief, thanks to its relational semantics, this can be done by looking at whether the operation that defines it preserves the corresponding relational properties. But bold distributed belief does not have relational semantics, so an identical approach will not work. Still, as it has been shown \autorefp{pro:DE:neigh:G}, $\dbold{}$ has neighbourhood semantics. This will be useful because the doxastic properties in question (truthfulness, consistency and introspection) can also be characterised by requirements on neighbourhood functions. \autoref{tb:property.neighbour.axiom} shows the details of this correspondence, with $\Box$ a standard `strict' neighbourhood modality \citep{ArecesF09} and $\Dia$ its dual. Proofs showing the correspondence can be found in \citet[Theorem 7.11]{Chellas1980}. A final warning: under relational semantics, seriality ($l$) is syntactically characterised not only by $\Box \varphi \rightarrow \Dia \varphi$, as indicated in \autoref{tb:property.axiom}, but also by $\lnot \Box \bot$. However, under neighbourhood semantics, $\Box \varphi \rightarrow \Dia \varphi$ and $\lnot \Box \bot$ are not equivalent: while the first ('no weak inconsistencies', because of its equivalent formulation as $\lnot (\Box \varphi \land \Box \lnot \varphi)$) corresponds to $l_N$, the second (`no strong inconsistencies') corresponds rather to $\emptyset \notin N(w)$. While those two formulas clearly indicate a form of consistency, here the focus will be in the first ($\Box \varphi \rightarrow \Dia \varphi$, and thus $l_N$). This is because $\lnot \dbold{G}{\bot}$ is valid (\autoref{pro:DE:absurdtautology} in \autoref{pro:DE}), and thus $\emptyset \notin N_G(w)$ is not possible.\label{con-weakstrong}

\begin{table}[t]
  \begin{footnotesizectabular}{l@{\qquad}l}
    \toprule
    \textbf{Frame condition}                                                      & \textbf{Characterising formula} \\
    \midrule
    $l_N$ (\emph{`seriality'}):                                                   & \emph{consistency}: \\
    $U \in N(w) \Rightarrow \comp{U} \notin N(w)$                            & $\Box \varphi \rightarrow \Dia \varphi$ \\
    \midrule
    $r_N$ (\emph{`reflexivity'}):                                                 & \emph{truthfulness of knowledge/belief}: \\
    $U \in N(w) \Rightarrow w \in U$                                              & $\Box \varphi \rightarrow \varphi$ \\
    \midrule
    $t_N$ (\emph{`transitivity'}):                                                & \emph{positive introspection}: \\
    $U \in N(w) \Rightarrow \set{w' \in W \mid U \in N(w')} \in N(w)$             & $\Box \varphi \rightarrow \Box \Box \varphi$ \\
    \midrule
    $s_N$ (\emph{`symmetry'})                                                     & \emph{truthfulness of possible knowledge/belief}: \\
    $w \in U \Rightarrow \set{w' \in U \mid \comp{U} \notin N(w')} \in N(w)$ & $\Dia \Box \varphi \rightarrow \varphi$ \\
    \midrule
    $e_N$ (\emph{`Euclidicity'}):                                                 & \emph{negative introspection}: \\
    $U \notin N(w) \Rightarrow \set{w' \in W \mid U \notin N(w')} \in N(w)$       & $\neg \Box \varphi \rightarrow \Box \neg \Box \varphi$ \\
    \bottomrule
  \end{footnotesizectabular}
  \caption{Neighbourhood properties and their characterising modal formula.}
  \label{tb:property.neighbour.axiom}
\end{table}

\smallskip


To make the connection precise, we start by giving neighbourhood semantics to the single-agent modality $\beli{}$.

\begin{prop}\label{pro:DE:neigh:a}
  Let $(\M, w)$ be a pointed belief model. For each $a \in A$, the define the \emph{neighbourhood function $N_a:W \to \wp(\wp(W))$} as $N_a(w) := \set{ U \subseteq W \mid \gcs{a}{w} \subseteq U}$. Then,
  \begin{center}
    \begin{tabular}[b]{l@{\qquad{iff}\qquad}l}
      $\M,w \vDash \beli{a}{\varphi}$ & $\extm{\varphi} \in N_a(w)$.\\
    \end{tabular}
  \end{center}
\end{prop}

Note: in non-serial models, $N_a$ \autorefp{pro:DE:neigh:a} and $N_{\set{a}}$ \autorefp{pro:DE:neigh:G} might be different (cf. \autoref{pro:DE:single.agent} in \autoref{pro:DE}).

\smallskip


Here is, then, the formal statement of the \emph{preservation} idea, now for a modality with neighbourhood semantics.

\begin{defi}[Preservation] \label{def:preserves.bold}
  Given a belief model $\M$ and a combination of neighbourhood frame conditions $\msf$ (i.e., $\msf \subseteq \set{l_N,r_N,t_N,s_N,e_N}$), we say that \emph{$\msf$ is preserved for bold distributed belief in $\M$} (alternatively, that \emph{bold distributed belief preserves $\msf$ in $\M$}) when, for any group $G \subseteq A$, the neighbourhood function $N_G$ \autorefp{pro:DE:neigh:G} satisfies $\msf$ whenever $N_a$ \autorefp{pro:DE:neigh:a} satisfies $\msf$ for every $a \in G$. A combination of neighbourhood frame conditions is preserved for bold distributed belief on a \emph{class of models} if and only if it is preserved in every model in that class.
\end{defi}

Finally, in the proof of the result \autorefp{pro:DE:rel-properties}, and when talking about the individual agents, it will be simpler to refer not to neighbourhood frame conditions but rather to relational frame conditions. We can do so due to the following well-known result (e.g., \citealp{Pacuit2017}), which establishes a correspondence between frame conditions on relational models and frame conditions in neighbourhood models.
It tells us that, for each of the five relational properties we consider, a belief model has that property if and only if the individual neighbourhood function has the corresponding property. So, for example, a belief model is reflexive (each individual agent's accessibility relation is reflexive) if and only if $N_a$ (for each agent $a \in A$, defined from that model) has the `neighbourhood-reflexivity' property $r_N$. When this is the case, even if we are talking about the neighbourhood functions defined on the model, we say that the model is `reflexive'.

\begin{prop}\label{prop:rel.neigh}
  Let $\M$ be the a belief model; take $a \in  A$. Let $x \in \set{l,r,t,s,e}$. Then, $R_a$ satisfies $x$ in $\M$ if and only if $N_a$ satisfies $x_N$ in $\M$.
\end{prop}


Here is, then, the result.

\begin{prop} \label{pro:DE:rel-properties}
  Take group $G \subseteq A$.\footnote{Note: different from the same result for cautious distributed belief (\autoref{pro:DA:rel-properties}), here the case with $\card{G}=1$ is still worthwhile to look at since, as mentioned, in non-serial models $N_a$ \autorefp{pro:DE:neigh:a} and $N_{\set{a}}$ \autorefp{pro:DE:neigh:G} might be different.}
  Then, bold distributed belief
  \begin{enumerate}
    \item 
    \begin{enumerate}[leftmargin=1.5em]
      \item\label{def:preserves-bold-1a} preserves $l_N$ on the class of reflexive models;
      \item does not preserve $l_N$ on the class of $\cond$ models, for any $\cond \subseteq \set{t,e}$;
      \item does not preserve $l_N$ on the class of symmetric models;
      \item preserves $l_N$ on the class of $\cond$ models, where $\cond \subseteq \set{s,t}$ or $\cond \subseteq \set{s,e}$;
    \end{enumerate}

    \item preserves $r_N$ on the class of all models;

    \item
    \begin{enumerate}[leftmargin=1.5em]
      \item does not preserve $t_N$ on the class of serial models;
      \item preserves $t_N$ on the class of reflexive models;
      \item preserves $t_N$ on the class of symmetric models;
      \item preserves $t_N$ on the class of Euclidean models;

    \end{enumerate}

    \item
    \begin{enumerate}[leftmargin=1.5em]
      \item does not preserve $s_N$ on the class of serial models;
      \item preserves $s_N$ on the class of reflexive models.
      \item does not preserve $s_N$ on the class of $\cond$ models, for any $\cond \subseteq \set{t,e}$;
      \item preserves $s_N$ on the class of $\cond$ models, when $\cond \subseteq \set{l,t}$ or $\cond \subseteq \set{l,e}$;
    \end{enumerate}

    \item
    \begin{enumerate}[leftmargin=1.5em]
      \item does not preserve $e_N$ on $\cond$ models, for any $\cond \subseteq \set{l,t}$;
      \item preserves $e_N$ on reflexive models;
      \item does not preserve $e_N$ on $\cond$ models, for any $\cond \subseteq \set{s,t}$;
      \item preserves $e_N$ on $\cond$ models for any $\set{l,s} \subseteq \cond$.
    \end{enumerate}
  \end{enumerate}
  \begin{proof}
    See the \hyperref[pro:DE:rel-properties:proof]{appendix}.
  \end{proof}
\end{prop}

As \autoref{pro:DA:rel-properties} did for $\dcaut{}{}$, \autoref{pro:DE:rel-properties} tells us, for each combination of the five frame conditions (on the relations), exactly which properties are preserved for $\dbold{}{}$ in each case (once again, the reader can refer to \autoref{tab:frameprops} in the appendix for an overview of equivalent combinations).

As opposed to $\dstan{}{}$ and $\dcaut{}{}$, the only frame condition preserved through $\dbold{}{}$ without additional assumptions is $r_N$. The frame condition $l_N$ (as well as $s_N$ and $e_N$) is preserved only under reflexivity, and thus bold distributed inherits lack of weak inconsistencies (as well as truthfulness of possible beliefs and negative introspection) from individual beliefs only when the different forms of distributed belief collapse.\footnote{Recall the discussion on page \pageref{con-weakstrong}: bold distributed beliefs always satisfies lack of strong inconsistencies.} Moreover, $t_N$ is preserved `more easily' for $\dbold{}{}$ than $t$ (transitivity) for $\dcaut{}{}$: while reflexivity or symmetry are enough for both, Euclidicity is also enough for $\dbold{}{}$. This means that, when the individual relations are assumed to be serial, transitive and Euclidean (so the individual belief modalities are KD45), bold distributed belief is positively introspective, unlike cautious distributed belief. Yet, bold distributed belief is not `proper' belief: it inherits neither lack of weak inconsistencies nor negative introspection in this class of models (besides not distributing over implications, as normal modalities do). In fact, the only classes of models where the properties discussed here are all preserved for bold distributed belief are the classes of models that are reflexive (and hence where all the discussed notions of distributed belief collapse).

\subsection{Relationship between \texorpdfstring{$\dbold{G}{}$}{DE} and \texorpdfstring{$\dstan{G}{}$}{D} and between \texorpdfstring{$\lang{\dbold{}{}}$}{DB} and \texorpdfstring{$\lang{\dstan{}{}}$}{D}}\label{sub:DEvsD}


Following the pattern of \autoref{sec:DA}, it is time to discuss the relationship between bold distributed belief and standard distributed belief. The proposition below establishes some initial connections between the modalities $\dbold{}$ and $\dstan{}$. First, $\dbold{}$ is definable in terms of $\dstan{}$ and Boolean operators, just like $\dcaut{}$. Second, bold distributed belief implies standard distributed belief, but not the other way around. In a picture,
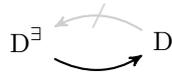
\begin{center}
  \begin{tikzpicture}[node distance = 2em and 3em, ->]
    \node[modality] (DE) {\dbold{}{}};
    \node[modality, right = of DE] (D) {\dstan{}{}};

    \path (DE) edge [bend right = 30pt] (D)
          (D)  edge [bend right = 30pt, verylightgray] node [sloped] {/} (DE);
  \end{tikzpicture}
\end{center}
Third, both notions (as well as $\dcaut{}$) coincide when the accessibility relations are reflexive.

\begin{prop} \label{pro:DE:defDG}
  ~
  \begin{enumerate}
    \item $\displaystyle \vDash \dbold{G}{\varphi} \leftrightarrow \bigvee_{\emptyset \neq G' \subseteq G} (\lnot \dstan{G'}{\bot} \land \dstan{G'}{\varphi})$.

    \item $\vDash \dbold{G}{\varphi} \limp \dstan{G}{\varphi}$ \; but \; $\not\vDash \dstan{G}{\varphi} \limp \dbold{G}{\varphi}$. \label{pro:DE:itm:str-D}

    \item Let \class{R} be the class of reflexive belief models. Then, $\class{R} \vDash \dbold{G}{\varphi} \leftrightarrow D_G \varphi$.
  \end{enumerate}
  \begin{proof}
    Take any model $\M = \langle W,R,\val \rangle$, any $w \in W$ and any group $G \subseteq A$.
    \begin{enumerate}
      \item {\prooflr} Suppose $\mps{w}\dbold{G}{\varphi}$; then, there is subgroup $G' \subseteq G$ such that $G'$ is (maximally) consistent at $w$ (so $\mps{w} \lnot \dstan{G'}{\bot}$) and it has a distributed belief in $\varphi$ (so $\mps{w}\dstan{G'}{\varphi}$). Hence, $\mps{w} \bigvee_{\emptyset\neq G' \subseteq G} (\lnot \dstan{G'}{\bot} \land \dstan{G'}{\varphi})$. {\proofrl} Suppose $\mps{w} \bigvee_{\emptyset \neq H \subseteq G} (\lnot \dstan{H}{\bot} \land \dstan{H}{\varphi})$, so there is subgroup $H \subseteq G$ that is consistent at $w$ and that has a distributed belief in $\varphi$. If $H$ is \emph{maximally} consistent at $w$ w.r.t. $G$, then $\mps{w}\dbold{G}{\varphi}$. Otherwise, some strict superset of $H$, say $G'$, is consistent and also maximally w.r.t. $G$ (that is, $H \subset G' \subseteq G$). Since $H$ has a distributed belief in $\varphi$ and $\dstan{}$ is group monotonic, then $G'$ also has a distributed belief in $\varphi$. Hence, $G'$ is a witness that makes $\mps{w}\dbold{G}{\varphi}$ true.

      \item For the first, $\mps{w} \dbold{G}{\varphi}$ implies there is $\maxatin{G'}{w}{G}$ such that $\mps{w} \dstan{G'}{\varphi}$. But $\dstan{}$ is group monotonic, so $\mps{w} \dstan{G}{\varphi}$. For the second, simply note that $\dstan{G}{\bot}$ is satisfiable but $\dbold{a}{\bot}$ is not.

      \item Immediate, as in reflexive models there is always exactly one maximally consistent subgroup of $G$: $G$ itself.
    \end{enumerate}
  \end{proof}
\end{prop}

Having shown the connection between $\dstan{}$ and $\dbold{}$, (the second part of \autoref{pro:DE:defDG}), it is time to compare their languages $\lang{\dstan{}{}}$ and $\lang{\dbold{}{}}$. Using the first part of \autoref{pro:DE:defDG}, one can define a translation that takes any formula in $\lang{\dbold{}{}}$ and returns a semantically equivalent formula in $\lang{\dstan{}{}}$.

\begin{coro}
  $\lang{\dstan{}{}}$ is at least as expressive as $\lang{\dbold{}{}}$ (in symbols: $\lang{\dbold{}{}} \preccurlyeq \lang{\dstan{}{}}$).
\end{coro}

For deciding whether the other direction ($\lang{\dstan{}{}} \preccurlyeq \lang{\dbold{}{}}$) holds, one can take advantage of the neighbourhood semantics of $\dbold{}{}$ and rely on bisimulations for such structures. Now, although \autoref{pro:DE:neigh:G} uses the `strict' semantics on a monotonic neighbourhood model, this is equivalent to the `loose' semantics on an arbitrary neighbourhood model. Thus, following \citet[Definition 2.2]{Pacuit2017} and \citet[Definition 4]{ArecesF09} (respectively), here is a first version of a bisimulation for $\dbold{}{}$ (called \emph{monotonic bisimulation} in \citealp{Pacuit2017}).

\begin{defi}[$\lang{\dbold{}{}}$-Bisimulation, v1] \label{def:DE:bisim}
  Let $\M=\langle W, R, \val \rangle$ and $\M'=\langle W',R',\val' \rangle$ be two belief models. A non-empty relation $Z \subseteq W \times W'$ is a \emph{$\lang{\dbold{}{}}$-bisimulation} between $\M$ and $\M'$ if and only if $Zww'$ implies all of the following.
  \begin{bisimitemize}
    \item \textbf{Atom.} For all $p \in P$: $w \in \val(p)$ if and only if $w' \in \val'(p)$.

    \item \textbf{Forth-1.} For all groups $G \subseteq A$ and all $\maxatin{H}{w}{G}$ there is a $\maxatin{H'}{w'}{G}$ such that, for all $u' \in W'$, if $u' \in \gcs{H'}{w'}$ then there is $u \in W$ such that $u \in \gcs{H}{w}$ and $Zuu'$.

    \item \textbf{Back-1.} For all groups $G \subseteq A$ and all $\maxatin{H'}{w'}{G}$ there is a $\maxatin{H}{w}{G}$ such that, for all $u \in W$, if $u \in \gcs{H}{w}$ then there is $u' \in W'$ such that $u' \in \gcs{H'}{w'}$ and $Zuu'$.
  \end{bisimitemize}
  Write $\M,w \bidboldover{Z} \M',w'$ when $Z$ is a $\lang{\dbold{}{}}$-bisimulation between $\M$ and $\M'$ with $Zww'$. Write $\M,w \bidbold \M',w$ when there is such a bisimulation $Z$.
\end{defi}

Note the quantification pattern of the \emph{forth-1} (resp., \emph{back-1}) clause:
\begin{smallctabular}{@{}c@{}}
  $\forall G \subseteq A \,.\, \forall \maxatin{H}{w}{G} \,.\, \exists \maxatin{H'}{w'}{G} \,.\, \forall u' \in W' \,.\, \left(
  \begin{array}{@{}c@{}}
    u' \in \gcs{H'}{w'} \\
    \Rightarrow \\
    \exists u \in W (u \in \gcs{H}{w} \;\&\; Zuu')
  \end{array}
  \right)$.
\end{smallctabular}
The pattern might look non-standard: leaving aside the quantification over groups of agents,\footnote{Recall: $G \neq \emptyset$.} the correspondence between worlds (every relevant world in $W'$ should have a matching world in $W$) is the opposite of what a standard `forth' clause does.

One can get back the `standard pattern' by instead working with \emph{sets} of worlds. Indeed, define, for $U \subseteq W$,
\[
  w \colleadsto{G'} U
  \quad\text{iff}_{\mathit{def}}\quad
  \gcs{G'}{w} = U
\]
so $w \colleadsto{G'} U$ indicates that, from $w$, the group $G'$ can collectively reach only worlds in $U$. Then, define
\[
  w \colmcssleadto{G} U
  \quad\text{iff}_{\mathit{def}}\quad
  w \colleadsto{G'} U \;\text{for some}\; \maxatin{G'}{w}{G}
\]
so $w \colmcssleadto{G} U$ indicates that some subgroup of $G$ that is maximally consistent at $w$ can collectively reach only worlds in $U$. This gives us an alternative definition for the semantic clause of $\dbold{}$: as the reader can check,
\begin{center}
  \begin{tabular}[b]{@{}l@{\quad{iff}\quad}l@{}}
    $\M,w \vDash \dbold{G}{\varphi}$ & $\exists U \subseteq \dom(\M)$ such that $w \colmcssleadto{G} U$ and $U \subseteq \extm{\varphi}$.
  \end{tabular}
\end{center}
Moreover, we can formulate the following alternative definition of an $\lang{\dbold{}{}}$-bisimulation.

\begin{defi}[$\lang{\dbold{}{}}$-Bisimulation, v2] \label{def:DE:bisim:bis}
  Let $\M=\langle W, R, \val \rangle$ and $\M'=\langle W',R',\val' \rangle$ be two belief models. A non-empty relation $Z \subseteq W \times W'$ is a \emph{$\lang{\dbold{}{}}$-bisimulation} between $\M$ and $\M'$ if and only if $Zww'$ implies all of the following.
  \begin{bisimitemize}
    \item \textbf{Atom.} As in \autoref{def:DE:bisim}.

    \item \textbf{Forth-2.} For all groups $G \subseteq A$ and all $U \subseteq W$: if $w \colmcssleadto{G} U$ then there is $U' \subseteq W'$ such that $w' \colmcssleadto{G} U'$ and $U' \subseteq Z[U]$, with
    \[
      \begin{array}{r@{\;:=\;}l}
        Z[U]       & \set{ u' \in W' \mid Zuu' \;\text{for some}\; u \in U} \\
      \end{array}
    \]

    \item \textbf{Back-2.} For all groups $G \subseteq A$ and all $U' \subseteq W'$: if $w' \colmcssleadto{G} U'$ then there is $U \subseteq W$ such that $w \colmcssleadto{G} U$ and $U \subseteq Z^{-1}[U']$, with
    \[
      \begin{array}{r@{\;:=\;}l}
        Z^{-1}[U'] & \set{ u \in W \mid Zuu' \;\text{for some}\; u' \in U'} \\
      \end{array}
    \]
  \end{bisimitemize}
\end{defi}

The quantification pattern in the \emph{forth-2} (resp., \emph{back-2}) clause, shown below, is closer to the one used by a standard `forth' condition.
\begin{smallctabular}{c}
  $\forall G \subseteq A \;.\; \forall T \subseteq W \;.\; \big(w \colmcssleadto{G} U \Rightarrow \exists U' \subseteq W (w' \colmcssleadto{G} U' \;\&\; U' \subseteq Z[U]) \big)$.
\end{smallctabular}
Probably more important: as the next proposition shows, the two formulations of $\lang{\dbold{}{}}$-bisimilarity are equivalent.

\begin{prop}\label{pro:bis1-bis2}
  Let $\M=\langle W, R, \val \rangle$ and $\M'=\langle W',R',\val' \rangle$ be two belief models. A non-empty relation $Z \subseteq W \times W'$ is a $\lang{\dbold{}{}}$-bisimulation according to \autoref{def:DE:bisim} if and only if it is a $\lang{\dbold{}{}}$-bisimulation according to \autoref{def:DE:bisim:bis}.
  \begin{proof}
    Take any non-empty $Z \subseteq W \times W'$.
    \begin{itemize}
      \item[\prooflr] Suppose $Z$ satisfies \autoref{def:DE:bisim}; it will be shown that it satisfies \autoref{def:DE:bisim:bis}. Take any $(w,w') \in W \times W'$ such that $Zww'$. Clause \textbf{atom} is automatic. For \textbf{forth-2}, take any group $G \subseteq A$ and any $U \subseteq W$; suppose $w \colmcssleadto{G} U$, that is, suppose $U = \gcs{H}{w}$ for some $\maxatin{H}{w}{G}$. Then, by \textbf{forth-1}, there is $\maxatin{H'}{w'}{G}$ such that, for all $u' \in W'$, if $u' \in \gcs{H'}{w'}$ then there is $u \in W$ such that $u \in \gcs{H}{w}$ and $Zuu'$. Define then $U' := \gcs{H'}{w'}$; it will be shown that $w' \colmcssleadto{G} U'$ and $U' \subseteq Z[U]$. The first follows from the definition of $U'$, as $\maxatin{H'}{w'}{G}$. For the second notice that, if $u' \in U' = \gcs{H'}{w'}$, then by \textbf{forth-1} there is $u \in W$ such that $u \in \gcs{H}{w} = U$ and $Zuu'$, and thus $u' \in Z[U]$. An analogous argument proves that \textbf{back-2} holds too.

      \item[\proofrl] Suppose $Z$ satisfies \autoref{def:DE:bisim:bis}; it will be shown that it satisfies \autoref{def:DE:bisim}. Take any $(w,w') \in W \times W'$ such that $Zww'$. Clause \textbf{atom} is automatic. For \textbf{forth-1}, take any group $G \subseteq A$ and any $\maxatin{H}{w}{G}$. Then, by definition, $w \colmcssleadto{G} \gcs{H}{w}$ so, by \textbf{forth-2}, there is $U' \subseteq W'$ such that $w' \colmcssleadto{G} U'$ and $U' \subseteq Z[\gcs{H}{w}]$. Now, from $w' \colmcssleadto{G} U'$ it follows that $U' = \gcs{H'}{w'}$ for some $\maxatin{H'}{w'}{G}$; it will be shown that this $H'$ is such that, for all $u' \in W'$, if $u' \in \gcs{H'}{w'}$ then there is $u \in W$ such that $u \in \gcs{H}{w}$ and $Zuu'$. So, take any $u' \in W'$ such that $u' \in \gcs{H'}{w'} = U'$. From $U' \subseteq Z[\gcs{H}{w}]$ it follows that $u' \in Z[\gcs{H}{w}]$, so there is $u \in W$ such that $u \in \gcs{H}{w}$ and $Zuu'$, as required. An analogous argument proves that \textbf{back-1} holds too.
    \end{itemize}
  \end{proof}
\end{prop}

Besides being equivalent, the alternative formulation makes it easier to check whether two belief models are $\lang{\dbold{}{}}$-bisimilar. Thus, it will be preferred through the rest of the text, with \textbf{forth-2} and \textbf{back-2} being referred to simply as \textbf{forth} for \textbf{back}.


\medskip

With this, we can put the notion of $\lang{\dbold{}{}}$-bisimilation to work.

\begin{defi}[$\lang{\dbold{}{}}$-equivalence]
  Two pointed models $\M,w$ and $\M',w'$ are $\lang{\dbold{}{}}$-equivalent (notation: $\M,w \eqdbold \M',w'$) if and only if, for every $\varphi \in \lang{\dbold{}{}}$,
  \begin{ctabular}{r@{\quad{if and only if}\quad}l}
    $\M, w \vDash \varphi$ & $\M',w' \vDash \varphi$.
  \end{ctabular}
  When the models are clear from context, we will write simply $w \eqdbold w'$.
\end{defi}

The following two propositions, consequences of \autoref{pro:bis1-bis2} together with results for monotonic bisimulation (\autoref{def:DE:bisim} here) in \citet{Pacuit2017} (Proposition 2.1 and Theorem 2.4, respectively), tell us that $\lang{\dbold{}{}}$-bisimulation characterizes modal equivalence for $\lang{\dbold{}{}}$ on individually image-finite models. The readers interested in direct proofs that use rather $\lang{\dbold{}{}}$-bisimulation as in \autoref{def:DE:bisim:bis} are referred to the appendix (\hyperref[pro:DE:invar:proof]{here} for the first and \hyperref[pro:DE:implies:proof]{here} for the second).

\begin{prop}[$\lang{\dbold{}{}}$-Bisimilarity implies $\lang{\dbold{}{}}$-equivalence]\label{pro:DE:invar}
  Let $\M,w$ and $\M',w'$ be pointed belief models. Then,
  \begin{ctabular}{r@{\quad{implies}\quad}l}
    $\M,w \bidbold \M',w'$ & $\M,w \eqdbold \M',w'$.
  \end{ctabular}
\end{prop}

\begin{prop}[$\lang{\dbold{}{}}$-Equivalence implies $\lang{\dbold{}{}}$-bisimilarity]\label{pro:DE:implies}
  Let $\M,w$ and $\M',w'$ be \emph{individually image-finite}\footnote{The proof in \citet[Theorem 2.4]{Pacuit2017} requires rather \emph{locally core-finite} neighbourhood models: those in which every neighbourhood $N(w)$ satisfies the following: \begin{inlineenum} \item it is core-complete (every $U \in N(w)$ is the superset of some subset-minimal element of $N(w)$), \item it is finite and \item $U \in N(w)$ implies $U$ is finite\end{inlineenum}. These three conditions are implied by the \emph{individually image-finite} requirement. The first because, as indicated in \citet[Page 3]{Pacuit2017}, this property holds when the neighbourhoods are monotonic (in our case, by definition in \autoref{pro:DE:neigh:G}) and only contain finitely many sets (in our case, because as $A$ is finite), the second by the finiteness of $A$, and the third by the individually image-finite requirement.} pointed belief models. Then,
  \begin{ctabular}{r@{\quad{implies}\quad}l}
    $\M,w \eqdbold \M',w'$ & $\M,w \bidbold \M',w'$.
  \end{ctabular}
\end{prop}

The models used in \autoref{fct:DA.less.D} where shown to be $\lang{\dcaut{}{}}$-bisimilar. As it turns out, the stated $\lang{\dcaut{}{}}$-bisimulation is also a $\lang{\dbold{}{}}$-bisimulation, allowing us to use them to show that $\lang{\dbold{}{}}$ is strictly less expressive than $\lang{\dstan{}{}}$.

\begin{fact}\label{fct:DE.less.D}
  $\lang{\dbold{}{}}$ is \emph{not} at least as expressive as $\lang{\dstan{}{}}$ (in symbols: $\lang{\dstan{}{}} \not\preccurlyeq \lang{\dbold{}{}}$).
  \begin{proof}
    Consider again the belief models from \autoref{fct:DA.less.D}:
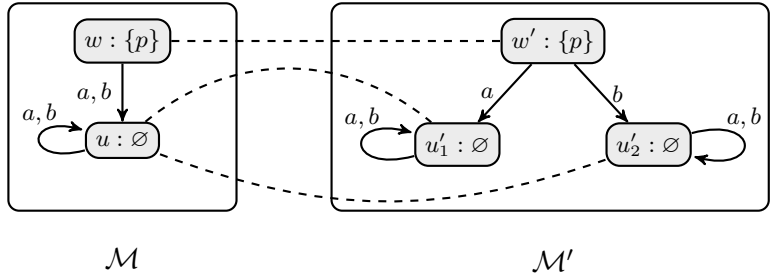
    \begin{center}
      \begin{tikzpicture}[node distance = 2em and 0em, ->]
        \node  at (0,0) [world] (w) {w:\set{p}};
        \node[world, below = of w] (u) {u:\emptyset};
        \path (w) edge node [left, label edge] {$a,b$} (u);
        \path (u) edge [loop left] node [above = 2pt, label edge] {$a,b$} ();
        \node [below = 6em of w] {$\M$};
        \draw [frame rectangle] (-1.5,-2.25) rectangle (1.5, 0.5);

        \node[world, right = 11.25em of w] (w') {w':\set{p}};
        \node[world, below left = of w'] (u'1) {u'_1:\emptyset};
        \node[world, below right = of w'] (u'2) {u'_2:\emptyset};
        \path (w') edge node [left, label edge] {$a$} (u'1)
                   edge node [right, label edge] {$b$} (u'2)
              (u'1) edge [loop left] node [above = 2pt, label edge] {$a,b$} ()
              (u'2) edge [loop right] node [above = 2pt, label edge] {$a,b$} ();
        \node [below = 6em of w'] {$\M'$};

        \path (w) edge [bisim] (w')
              (u) edge [bisim, bend left = 40pt] (u'1)
                  edge [bisim, bend right = 20pt] (u'2);
        \draw [frame rectangle] (2.75,-2.25) rectangle (8.5, 0.5);
      \end{tikzpicture}
    \end{center}  
    The dashed edges (same as before) also define a $\lang{\dbold{}{}}$-bisimulation between $\M$ and $\M'$, as the pairs $(w,w')$, $(u,u'_1)$ and $(u,u'_2)$ satisfy the three clauses. The proof follows a similar pattern to that of fact \autoref{fct:DA.less.D}, using $\maxcsat{G}{v}$ to denote the set of conjecture sets of subgroups of $G$ that are maximally consistent at a world $v$ (i.e., $\maxcsat{G}{v}:=\set{U\subseteq W \mid v \colmcssleadto{G}U}$). 
    \begin{compactitemize}
      \item $\bm{(w,w')}$. The \textbf{atom} clause is immediate. \textbf{Forth-2}. First, consider $G=\set{a}$. Then, $\maxcsat{\set{a}}{w}= \set{\set{u}}$, while $\maxcsat{\set{a}}{w'}=\set{\set{u'_1}}$, and $Z[\set{u}]=\set{u'_1,u'_2}\supseteq \set{u'_1}$. Similarly for $G=\set{b}$. For $G=\set{a,b}$, $\maxcsat{G}{w}=\set{\set{u}}$, while $\maxcsat{G}{w'}= \set{\set{u'_1},\set{u_2'}}$. But, $Z[\set{u}]=\set{u'_1,u'_2}$, and so either element of $\maxcsat{G}{w'}$ is sufficient to satisfy \textbf{forth-2}. The \textbf{back-2} clause follows a similar pattern, where for $G=\set{a,b}$: $Z^{-1}[\set{u'_1}]=Z^{-1}[\set{u'_2}]=\set{u}$, and $\set{u} \in \maxcsat{G}{w}$.

      \item $\bm{(u,u_1')}$. $\maxcsat{\set{a}}{u}=\maxcsat{\set{b}}{u}=\maxcsat{\set{a,b}}{u}=\set{u}$, and $\maxcsat{\set{a}}{u'_1}=\maxcsat{\set{b}}{u'_1}=\maxcsat{\set{a,b}}{u'_1}=\set{u'_1}$. \textbf{Forth-2} then follows from $Z[\set{u}]=\set{u'_1,u'_2} \supseteq \set{u'_1}$ and \textbf{back-2} from $Z^{-1}[\set{u'_1}]=\set{u}$.
      
      \item $\bm{(u,u_2')}$. As the previous case.
    \end{compactitemize}
    Thus, $M,w \bidbold \M',w'$ and hence $\M,w \eqdbold \M',w'$. However, the pointed models can be distinguished by a formula in $\lang{\dstan{}{}}$, as $\mpf{w} \dstan{\set{a,b}}{\bot}$ and yet $\M',w' \vDash \dstan{\set{a,b}}{\bot}$. Therefore $\lang{\dstan{}{}} \not\preccurlyeq \lang{\dbold{}{}}$.
  \end{proof}
\end{fact}

\begin{coro}\label{coro:DE.less.D}
  $\lang{\dstan{}{}}$ is strictly more expressive than $\lang{\dbold{}{}}$ (symbols: $\lang{\dbold{}{}} \prec \lang{\dstan{}{}}$).
\end{coro}

Thus, just like $\lang{\dcaut{}{}}$, the language $\lang{\dbold{}{}}$ is strictly less expressive than $\lang{\dstan{}{}}$. Even more: just like $\lang{\dcaut{}{}}$, adding the group inconsistency constant $\inc{G}$ is sufficient for $\lang{\dbold{}{}}$ to become as expressive as $\lang{\dstan{}{}}$.

\begin{prop}\label{prp:equal:DE:D}
  $\lang{\dbold{}{}, \inc{}}$ and $\lang{\dstan{}{}}$ are equally expressive (symbols: $\lang{\dbold{}{}, \inc{}} \approx \lang{\dstan{}{}}$).
  \begin{proof}
    Similar to that of \autoref{prp:equal}.
    First, $\vDash \dstan{G}{\varphi} \ldimp (\inc{G} \lor \dbold{G}{\varphi})$, so $\dstan{G}{}$ is definable in $\lang{\dbold{}{}, \inc{}}$. To see that the validity holds: from left to right, assume $\mps{w} \dstan{G}{\varphi}$. Now, if $\gcs{G}{w} \neq \emptyset$ then, since $\forall w'\in \gcs{G}{w}$, $\mps{w'}\varphi$, it follows that $\mps{w}\dbold{G}{\varphi}$. Otherwise, $\gcs{G}{w}=\emptyset$ and then $\mps{w}\inc{G}$. From right to left, just note that $\vDash \dbold{G}{\varphi}\limp \dstan{G}{\varphi}$ (\autoref{pro:DE:itm:str-D} in \autoref{pro:DE}) and $\vDash \inc{G} \rightarrow \dstan{G}{\varphi}$ for an arbitrary $\varphi$.
  \end{proof}
\end{prop}

\subsection{Relationship between \texorpdfstring{$\dbold{G}{}$}{DE} and \texorpdfstring{$\dcaut{G}{}$}{DA} and between \texorpdfstring{$\lang{\dbold{}{}}$}{DE} and \texorpdfstring{$\lang{\dcaut{}{}}$}{DA}}\label{sub:DEvsDA}

Having looked at the relationship between $\dstan{}{}$ and both $\dbold{}{}$ and $\dcaut{}{}$, the thing left to do is look at the relationship between $\dcaut{}{}$ and $\dbold{}{}$. The proposition below establishes some connection between these two modalities. First, $\dbold{}{}$ is definable in terms of $\dcaut{}{}$ and Boolean operators. Second, when considering arbitrary models, the modalities are independent (neither implies the other). In a picture,
\begin{center}
  \begin{tikzpicture}[node distance = 2em and 3em, ->]
    \node[modality] (DA) {\dcaut{}{}};
    \node[modality, below = of DE] (DE) {\dbold{}{}};

    \path (DA) edge [bend left = 30pt, verylightgray] node [sloped] {/} (DE)
          (DE) edge [bend left = 30pt, verylightgray] node [sloped] {/} (DA);
  \end{tikzpicture}
\end{center}
Third: in serial models, $\dcaut{}{}$ implies $\dbold{}{}$.

\begin{prop}\label{pro:DE:DA}
  ~
  \begin{enumerate}    
    \item\label{pro:DE:DA:itm:1} $\displaystyle \vDash \dbold{G}{\varphi} \ldimp \bigvee_{G' \subseteq G}(\dcaut{G'}{\varphi} \land \neg \dcaut{G'}{\bot})$.
    
    \item $\nvDash \dcaut{G}{\varphi} \limp \dbold{G}{\varphi}$ \;and\; $\nvDash \dbold{G}{\varphi} \limp \dcaut{G}{\varphi}$.

    \item Let $\class{L}$ be the class of serial belief models. Then, $\class{L} \vDash \dcaut{G}{\varphi} \limp \dbold{G}{\varphi}$ (and yet $\class{L} \nvDash \dbold{G}{\varphi} \limp \dcaut{G}{\varphi}$).
  \end{enumerate}
  \begin{proof}
    ~
    \begin{enumerate}      
      \item $\mps{s} \dbold{G}{\varphi}$ if and only if there is some $\maxatin{G'}{s}{G}$ such that $\forall s' \in \gcs{G'}{s}$, $\mps{s'} \varphi$, if and only if $\mps{s} \dcaut{G'}{\varphi} \land \lnot \dcaut{G'}{\bot}$. Thus, $\mps{s} \dbold{G}{\varphi}$ if and only if $\mps{s} \bigvee_{G' \subseteq G} (\dcaut{G'}{\varphi} \land \lnot \dcaut{G'}{\bot})$.      

      \item For the first, recall that $\vDash \dcaut{G}{\top}$ but $\nvDash \dbold{G}{\top}$. For the second, just recall that a group can have disjoint maximally consistent subgroups, and thus their respective combined conjecture sets might be disjoint too.
      
      \item For the first note that, in serial models, every group $G$ has at least one maximally consistent subgroup at any world $w$. Thus, assuming the universal quantification in $\dcaut{G}{\varphi}$ implies the existential quantification in $\dbold{G}{\varphi}$. For the second, the second argument in the previous item holds for serial models too.
    \end{enumerate}
  \end{proof}
\end{prop}

For the purposes of comparing the languages using the modalities $\dbold{}{}$ and $\dcaut{}{}$, \autoref{pro:DE:DA:itm:1} just above tells us that there is a translation taking formulas in $\lang{\dbold{}{}}$ and returning semantically equivalent formulas in $\lang{\dcaut{}{}}$. Thus, the latter is at least as expressive as the former.

\begin{coro}
  $\lang{\dcaut{}{}}$ is at least as expressive as $\lang{\dbold{}{}}$ (in symbols: $\lang{\dbold{}{}} \preccurlyeq \lang{\dcaut{}{}}$).
\end{coro}

To show that the other direction fails (i.e., $\lang{\dcaut{}{}} \not\preccurlyeq \lang{\dbold{}{}}$), bisimulation can be used once again. What is needed now is two models that are $\lang{\dbold{}{}}$-bisimilar but can be distinguished by $\lang{\dcaut{}{}}$.

\begin{fact}\label{fct:DE.less.DA.alt}
  $\lang{\dbold{}{}}$ is \emph{not} at least as expressive as $\lang{\dcaut{}{}}$ (in symbols: $\lang{\dcaut{}{}} \not\preccurlyeq \lang{\dbold{}{}}$).
  \begin{proof}
    Consider the belief models below. 
    
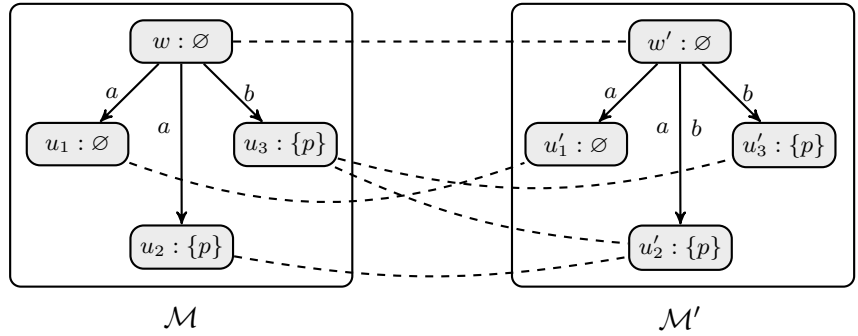
\begin{center}
      \begin{tikzpicture}[node distance = 2em and 0em, ->]
        \node  at (0,0) [world, minimum size = 16pt, minimum width = 38pt] (w) {w:\emptyset};
        \node[world, minimum size= 16pt, minimum width = 38pt, below left = of w] (u1) {u_1:\emptyset};
        \node[world, minimum size= 16pt, minimum width = 38pt, below right = of u1] (u2) {u_2:\set{p}};
        \node[world, minimum size= 16pt, minimum width = 38pt, below right = of w] (u3) {u_3:\set{p}};
        \path (w) edge node [left, label edge] {$a$} (u1)
                  edge node [left, label edge, pos=0.4] {$a$} (u2)
                  edge node [right, label edge] {$b$} (u3);
        \node [below = 8em of w] {$\M$};

        \node[world, minimum size = 16pt, minimum width = 38pt, right = 13.5em of w] (w') {w':\emptyset};
        \node[world, minimum size = 16pt, minimum width = 38pt, below left = of w'] (u'1) {u'_1:\emptyset};
        \node[world, minimum size = 16pt, minimum width = 38pt, below right =of u'1] (u'2) {u'_2:\set{p}};
        \node[world, minimum size = 16pt, minimum width = 38pt, below right = of w'] (u'3) {u'_3:\set{p}};
        \path (w') edge node [left, label edge] {$a$} (u'1)
                   edge node [left, label edge, pos = 0.4] {$a$} node [right, label edge, pos = 0.4] {$b$} (u'2)
                   edge node [right, label edge] {$b$} (u'3);
        \node [below = 8em of w'] {$\M'$};

        \path (w) edge [bisim] (w')
              (u1) edge [bisim, bend right = 20pt] (u'1)
              (u2) edge [bisim, bend right = 10pt] (u'2)
              (u3) edge [bisim, bend right = 15pt] (u'3)
                  edge [bisim, bend right =10pt] (u'2);
        \draw [frame rectangle] (4.35,-3.25) rectangle (8.85, 0.5);
        \draw [frame rectangle] (-2.25,-3.25) rectangle (2.25, 0.5);
      \end{tikzpicture}
    \end{center}
    The dashed edges define a $\lang{\dbold{}{}}$-bisimulation between $\M$ and $\M'$, as the pairs $(w,w')$, $(u_1,u'_1)$, $(u_2,u'_2)$, $(u_3,u_2')$ and $(u_3,u_3')$ satisfy the three clauses. In each case, the \textbf{atom} clause is immediate. For all pairs except $(w,w')$, the \textbf{forth} and \textbf{back} conditions are vacuously satisfied (as there are no consistent sets of agents at any of the relevant worlds in either model).
    For $(w,w')$, use again $\maxcsat{G}{v}$ to denote the set of conjecture sets of subgroups of $G$ that are maximally consistent at a world $v$ ($\maxcsat{G}{v}:=\set{U\subseteq W \mid v \colmcssleadto{G}U}$). First, take $G= \set{a}$; then, $\maxcsat{G}{w}= \set{V{=}\set{u_1,u_2}}$ and $\maxcsat{G}{w'}=\set{V'{=}\set{u_1',u_2'}}$. For \textbf{forth}, note that for $V \in \maxcsat{G}{w}$ there is $V' \in \maxcsat{G}{w'}$ with $V' \subseteq Z[V] = \set{u'_1,u'_2}$; for \textbf{back}, note that for $V' \in \maxcsat{G}{w'}$ there is $V \in \maxcsat{G}{w}$ with $V \subseteq Z^{-1}[V'] = \set{u_1,u_2,u_3}$.
    Now take $G= \set{b}$, so $\maxcsat{G}{w}= \set{V{=}\set{u_3}}$ and $\maxcsat{G}{w'}=\set{V'{=}\set{u_2',u_3'}}$. For \textbf{forth}, note that for $V \in \maxcsat{G}{w}$ there is $V' \in \maxcsat{G}{w'}$ with $V' \subseteq Z[V] = \set{u'_2,u'_3}$; for \textbf{back}, note that for $V' \in \maxcsat{G}{w'}$ there is $V \in \maxcsat{G}{w}$ with $V \subseteq Z^{-1}[V'] = \set{u_2,u_3}$. Finally, take $G = \set{a,b}$, so $\maxcsat{G}{w}=\set{V_1{=}\set{u_1,u_2}, V_2{=}\set{u_3}}$ and $\maxcsat{G}{w'}=\set{V'{=}\set{u_2'}}$. For \textbf{forth} note that, on the one hand, for $V_1 \in \maxcsat{G}{w}$ there is $V' \in \maxcsat{G}{w'}$ with $V' \subseteq Z[V_1] = \set{u'_1,u'_2}$, and on the other hand, for $V_2 \in \maxcsat{G}{w}$ there is $V' \in \maxcsat{G}{w'}$ with $V' \subseteq Z[V_2] = \set{u'_2,u'_3}$. For \textbf{back}, note that for $V' \in \maxcsat{G}{w'}$ there is $V_2 \in \maxcsat{G}{w}$ with $V_2 \subseteq Z^{-1}[V'] = \set{u_2,u_3}$. Thus, the dashed edges define a $\lang{\dbold{}{}}$-bisimulation, so $M,w \bidbold \M',w'$ and hence $\M,w \eqdbold \M',w'$.

    However, these pointed models can be distinguished by a formula in $\lang{\dcaut{}{}}$, as $\mpf{w} \dcaut{\set{a,b}}{p}$ and yet $\M',w' \vDash \dcaut{\set{a,b}}{p}$. Therefore $\lang{\dcaut{}{}} \not\preccurlyeq \lang{\dbold{}{}}$.
  \end{proof}
\end{fact}



\begin{coro}
  $\lang{\dcaut{}{}}$ is strictly more expressive than $\lang{\dbold{}{}}$ (symbols: $\lang{\dbold{}{}} \prec \lang{\dcaut{}{}}$).\footnote{Note: this, together with \autoref{coro:DA.less.D}, imply already the result stated in \autoref{coro:DE.less.D}.}
\end{coro}

\section{Summary and further work} \label{sec:end}

This paper has presented and discussed two forms of distributed belief, \emph{cautious} and \emph{bold}, aimed to capture a form of (group) distributed belief that does not collapse in the face of conflicting information. Both notions work by considering maximally consistent subgroups: while a set of agents $G$ has \emph{cautious} distributed belief that $\varphi$ ($\dcaut{G}{\varphi}$) if and only if $\varphi$ is true in every world in the conjecture set of \emph{every} maximally consistent subgroup of $G$, the group has \emph{bold} distributed belief that $\varphi$ ($\dbold{G}{\varphi}$) if and only if $\varphi$ is true in every world in the conjecture set of \emph{some} maximally consistent subgroup of $G$. As a result, while the cautious distributed belief of a group is inconsistent only when all the members of the group are individually inconsistent, the bold distributed belief of a group is never inconsistent. Thus, the introduced notions `behave better' than standard distributed belief, which can be inconsistent even when none of the members of the group are.

\medskip

After presenting these notions' respective modalities and semantic interpretations, the paper discussed some of their basic properties. For the \emph{cautious} notion, its singleton case $\dcaut{\set{a}}{}$ is equivalent to individual belief $\beli{a}{}$ and it lacks group monotonicity. Maybe more importantly, it has relational semantics, and thus it is a normal modality. For the \emph{bold} notion, its singleton case $\dbold{\set{a}}{}$ is \emph{not} equivalent to individual belief $\beli{a}{}$ (but $\beli{a}{}$ can be defined with $\dbold{\set{a}}{}$ and Boolean operators) and it has group monotonicity. Maybe more importantly, while it does not have relational semantics, it has neighbourhood semantics, satisfying additionally closure under conjunction elimination but lacking necessitation and closure under conjunction introduction.

\smallskip

The paper then studied to which extent a given property of individual belief `carries over' to each one of the introduced notions. For the \emph{cautious} notion, truthfulness and consistency carry over without additional assumptions. Moreover, while truthfulness of possible belief and negative introspection are both preserved in the presence of truthfulness, positive introspection is preserved in the presence of either truthfulness or truthfulness of possible beliefs. For the \emph{bold} notion, while truthfulness carries over on its own and positive introspection carries over with the help of truthfulness, truthfulness of possible beliefs or negative introspection, the three other properties (lack of weak inconsistencies, truthfulness of possible beliefs and negative introspection) are preserved only under truthfulness.

\smallskip

Finally, the paper compared standard, cautious and bold distributed belief. At the level of modalities, and in arbitrary belief models, cautious and bold distributed belief are independent of each other, and yet both imply standard distributed belief, as summarised in the following picture.
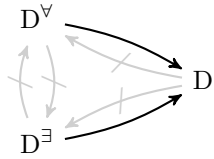
\begin{center}
  \begin{tikzpicture}[node distance = 0.5em and 4em, ->]
    \node[modality] (D) {\dstan{}{}};
    \node[modality, above left = of D] (DA) {\dcaut{}{}};
    \node[modality, below left = of D] (DE) {\dbold{}{}};

    \path (DA) edge [bend left = 20pt, verylightgray] node [sloped] {/} (DE)
               edge [bend left = 10pt] (D)
          (DE) edge [bend left = 20pt, verylightgray] node [sloped] {/} (DA)
               edge [bend right = 10pt] (D)
          (D)  edge [bend left = 10pt, verylightgray] node [sloped] {/} (DA)
               edge [bend right = 10pt, verylightgray] node [sloped] {/} (DE);
  \end{tikzpicture}
\end{center}
Then, at the level of languages, the expressivity of the languages forms a strict linear order: 
\begin{center}
  $\lang{\dbold{}{}} \prec \lang{\dcaut{}{}} \prec \lang{\dstan{}{}}$
\end{center}

\medskip

Among the questions that still need answer, the main ones are an axiom system for both languages $\lang{\dcaut{}{}}$ and $\lang{\dbold{}{}}$. Given the definability results, the modalities $\dcaut{}{}$ and $\dbold{}{}$ are not needed in the presence of $\dstan{}{}$ (or, from a different perspective, the language $\lang{\dstan{}{}, \dcaut{}{}, \dbold{}{}}$ is axiomatised by an axiom system for $\lang{\dstan{}{}}$ plus the validities $\dcaut{G}{\varphi} \leftrightarrow \bigwedge_{\emptyset \neq G' \subseteq G} \Big( \big( \lnot D_{G'} \bot \land \bigwedge_{G' \subset H \subseteq G} D_{H} \bot \big) \rightarrow D_{G'}\varphi \Big)$ and $\dbold{G}{\varphi} \leftrightarrow \bigvee_{\emptyset \neq H \subseteq G} (\lnot \dstan{H}{\bot} \land \dstan{H}{\varphi})$. However, one would like to find an axiom system that characterises $\dcaut{}{}$ and $\dbold{}{}$ in their own terms. Another interesting question is the languages' complexity profile. Then, among the further research lines, the natural one is to look into explicit epistemic actions that, in the style of \citet{agotnes20171,BaltagS20}, `resolve' the introduced forms of distributed belief.

\appendix
\section{Appendix}\label{sec:appendix}

\subsection*{Frame conditions}

\begin{table}[t!]
  \begin{center}
    \begin{footnotesize}
      \begin{tabular}{lr}
        \toprule
        \textbf{Frame conditions} & \textbf{Generic name} \\
        \midrule
        ---&K\\
        l&D\\
        r, lr&T\\
        t&K4\\
        s&KB\\
        e&K5\\
        lt&KD4\\
        ls&KDB\\
        le&KD5\\
        rt, lrt&S4\\
        rs, lrs&B\\
        re, lre, lts, lse, rts, rte, rse, lrts, lrte, lrse, ltse, rtse, lrtse&S5\\
        ts, se, tse&K4B\\
        te&K45\\
        lte& KD45 \\
        \bottomrule
      \end{tabular}
    \end{footnotesize}
  \end{center}
  \caption{Combinations of frame conditions and their generic name.}
  \label{tab:frameprops}
\end{table}

\autoref{tab:frameprops} shows the 32 combinations of frame conditions, and the generic names of the corresponding logics. Combinations on the same line are equivalent.

\subsection*{Proof of \autoref{pro:DA:rel-properties}}\label{pro:DA:rel-properties:proof}

\begin{figure}[t]
  \begin{ctabular}{@{}c@{}}
    \begin{tabular}{c@{\quad}c@{\quad}c}
      \begin{subfigure}[b]{0.3\textwidth}
        \centering
        \begin{tikzpicture}[frame rectangle, node distance = 2.5em and 1.5em, ->]
          \node[world] (w1) {w_1};
          \node[world, right = of w1] (w2) {w_2};
          \node[world, right = of w2] (w3) {w_3};
          \path (w1) edge [loop above] node [above = 2pt, label edge] {$b$} ()
                     edge node [above, label edge] {$a$} (w2)
                (w2) edge [loop above] node [above = 2pt, label edge] {$a$} ()
                     edge node [above, label edge] {$b$} (w3)
                (w3) edge [loop above] node [above = 2pt, label edge] {$a,b$} ();
        \end{tikzpicture}
        \caption{$F_1$}
        \label{fig:f1}
      \end{subfigure}
      &
      \begin{subfigure}[b]{0.25\textwidth}
        \centering
        \begin{tikzpicture}[frame rectangle, node distance = 2.5em and 1.5em, ->]
          \node[world] (w) {w_1};
          \node[world, below = of w] (u) {w_2};
          \path (w) edge[<->] node [left, label edge] {$a$} (u)
                (w) edge [loop left] node [above = 2pt, label edge] {$a$} ()
                (u) edge [loop left] node [above = 2pt, label edge] {$a,b$} ();
        \end{tikzpicture}
        \caption{$F_2$}
        \label{fig:f2}
      \end{subfigure}
      &
      \begin{subfigure}[b]{0.3\textwidth}
        \centering
        \begin{tikzpicture}[frame rectangle, node distance = 2.5em and 1.5em, ->]
          \node[world] (w) {w_1};
          \node[world, left = of w] (u1) {w_2};
          \node[world, right = of w] (u2) {w_3};
          \path (w)
              edge[<->] node [above, label edge] {$a$} (u1)
              edge[<->] node [above, label edge] {$b$} (u2)
            (u1) edge [loop above] node [above = 2pt, label edge] {$a,b$} ()
            (u2) edge [loop above] node [above = 2pt, label edge] {$a,b$} ();
        \end{tikzpicture}
        \caption{$F_3$}
        \label{fig:f13}
      \end{subfigure}
    \end{tabular}
    \\[1em]
    \begin{tabular}{c@{\qquad\qquad}c}
      \begin{subfigure}[b]{0.35\textwidth}
        \centering
        \begin{tikzpicture}[frame rectangle, node distance = 2.5em and 0.5em, ->]
          \node[world] (w) {w_1};
          \node[world, below left = of w] (u) {w_2};
          \node[world, below right = of w] (u2) {w_3};
          \path (w) edge node [left, label edge] {$a$} (u)
                (u2) edge [loop right] node [above = 2pt, label edge] {$a,b$} ()
                (u) edge [loop left] node [above = 2pt, label edge] {$a,b$} ()
                (w) edge node [right, label edge] {$b$} (u2);
        \end{tikzpicture}
        \caption{$F_4$}
        \label{fig:f4}
      \end{subfigure}
      &
      \begin{subfigure}[b]{0.35\textwidth}
        \centering
        \begin{tikzpicture}[frame rectangle, node distance = 2.5em and 0.5em, ->]
          \node[world, right = 12em of w] (w') {w_1};
          \node[world, below left = of w'] (u'1) {w_2};
          \node[world, below right = of w'] (u'2) {w_3};
          \path (w') edge[<->] node [left, label edge] {$a$} (u'1)
                     edge[<->] node [right, label edge] {$a$} (u'2)
                (w') edge [loop left] node [above = 2pt, label edge] {$a$} ()
               (u'1) edge [loop left] node [above = 2pt, label edge] {$a,b$} ()
               (u'2) edge [loop right] node [above = 2pt, label edge] {$a$} ()
               (u'1) edge[<->] node [above = 2pt, label edge] {$a$} (u'2);
        \end{tikzpicture}
        \caption{$F_5$}
        \label{fig:f5}
      \end{subfigure}
    \end{tabular}
  \end{ctabular}
  \caption{Counterexamples used in the proof of \autoref{pro:DA:rel-properties}}
  \label{tbl:counterexamples}
\end{figure}
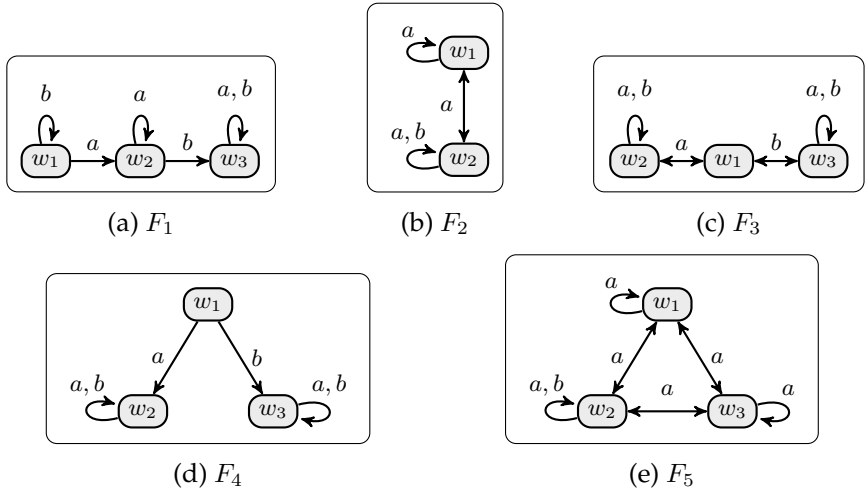

For the positive results, let $\M = \langle W,R,\val \rangle$ be a model.
\begin{enumerate}
  \item Suppose $\M$ is serial; take $w \in W$. Recall that $G \neq \emptyset$, so there is $a \in G$; by seriality, $\gcs{a}{w} \neq \emptyset$, that is, $a$ is consistent at $w$. Thus, there is at least one $G' \subseteq G$ that is maximally consistent at $w$ (one containing $a$). Hence, there is $u \in \gcs{G'}{w} \subseteq \cgcs{G}{w}$, as seriality requires.

  \item Suppose $\M$ is reflexive; take $w \in W$. By reflexivity, $w \in \gcs{a}{w}$ for every $a \in G$, so $G$ itself is its own single maximally consistent subset. Then, $\gcs{G}{w} = \cgcs{G}{w}$ so $w \in \cgcs{G}{w}$ and thus $w\crel{G}w$, as reflexivity requires.

  \item
  \begin{enumerate}[leftmargin=1.5em]
    \item For this negative result, consider frame $F_1$ in \autoref{fig:f1}. Relations $R_a$ and $R_b$ are transitive, serial and Euclidean. Still, $R^\forall_{\set{a,b}} \!=\! \set{(w_1, w_1), (w_1, w_2), (w_2, w_2), (w_2, w_3), (w_3, w_3)}$ is not transitive.

    \item Suppose $\M$ is reflexive; take $w, u, v \in W$ such that $w \crel{G} u$ and $u \crel{G} v$, and assume $a \in G$ implies $R_a$ is transitive. By reflexivity, $G$ is the only maximally consistent subset at both $w$ and $u$, so $\gcs{G}{w} = \cgcs{G}{w}$ and $\gcs{G}{u} = \cgcs{G}{u}$. Then, $wR_au$ and $uR_av$ for every $a \in G$, which by transitivity implies $wR_av$ for all such $a$. Thus, $v \in \cgcs{G}{w}$ and hence $w\crel{G}v$, as transitivity requires.

    \item Suppose $\M$ is symmetric; take $w, u, v \in W$ such that $w \crel{G} u$ and $u \crel{G} v$, and assume $a \in G$ implies $R_a$ is transitive. Then, there are $\maxatin{H_1}{w}{G}$ and $\maxatin{H_2}{u}{G}$ such that $u \in \gcs{H_1}{w}$ and $v \in \gcs{H_2}{u}$. By symmetry, $w \in \gcs{H_1}{u}$ and $u \in \gcs{H_2}{v}$; then, by transitivity, $u \in \gcs{H_1}{u}$ and $u \in \gcs{H_2}{u}$. But then, $H_1 \cup H_2$ is consistent at $u$ and, since $H_2$ is maximally consistent at $u$, then $(H_1 \cup H_2) \subseteq H_2$, that is, $H_1 \subseteq H_2$. Hence, the previous $v \in \gcs{H_2}{u}$ implies $v \in \gcs{H_1}{u}$ which, together with $u \in \gcs{H_1}{w}$ and transitivity implies $v \in \gcs{H_1}{w}$. Finally, $v \in \cgcs{G}{w}$ (since $H_1$ is maximally consistent at $w$ w.r.t. $G$) and hence $w \crel{G} v$, as transitivity requires.
  \end{enumerate}

  \item
  \begin{enumerate}[leftmargin=1.5em]
    \item For this negative result, consider frame $F_2$ in \autoref{fig:f2}. Relations $R_a$ and $R_b$ are symmetric, transitive and Euclidean. Still, $R^\forall_{\set{a,b}} = \set{(w_1, w_1), (w_1, w_2), (w_2, w_2)}$ is not symmetric.

    \item For this negative result, consider frame $F_3$ in \autoref{fig:f13}. Relations $R_a$ and $R_b$ are symmetric and serial. Still, the relation $R^\forall_{\set{a,b}} = \set{(w_1, w_2), (w_1, w_3), (w_2, w_2), (w_3, w_3)}$ is not symmetric. 

    \item Symmetry, seriality and Euclidicity together imply reflexivity, so \autoref{pro:DA:rel-properties:sym:ref} just below gives us the required result.

    \item\label{pro:DA:rel-properties:sym:ref} Suppose $\M$ is reflexive; take $w, u \in W$ such that $w \crel{G} u$, and assume $a \in G$ implies $R_a$ is symmetric. By reflexivity, $G$ is the only maximally consistent subset at both $w$ and $u$, so $\gcs{G}{w} = \cgcs{G}{w}$ and $\gcs{G}{u} = \cgcs{G}{u}$. Then, $wR_au$ for every $a \in G$, which by symmetry implies $uR_aw$ for all such $a$. Thus, $w \in \cgcs{G}{u}$ and hence $u\crel{G}w$, as required by symmetry.
  \end{enumerate}

  \item
  \begin{enumerate}[leftmargin=1.5em]
    \item Again, symmetry, seriality and Euclidicity imply reflexivity, so \autoref{pro:DA:rel-properties:euc:ref} below gives us the required result.

    \item For this negative result, consider frame $F_4$ in \autorefp{fig:f4}. Relations $R_a$ and $R_b$ are Euclidean, serial and transitive. Still, $R^\forall_{\set{a,b}} = \set{(w_1, w_2),  (w_1, w_3), (w_2, w_2), (w_3, w_3)}$ is not Euclidean.

    \item For this negative result, consider frame $F_5$ in \autorefp{fig:f5}. Relations $R_a$ and $R_b$ are Euclidean, symmetric and transitive. Still, $R^\forall_{\set{a,b}} = (W \times W) \setminus \set{(w_2, w_1), (w_2, w_3)}$ is not Euclidean.
    
    \item\label{pro:DA:rel-properties:euc:ref} Suppose $\M$ is reflexive; take $w, u, v \in W$ such that $w \crel{G} u$ and $w \crel{G} v$, and assume $a \in G$ implies $R_a$ is Euclidean. By reflexivity, $G$ is the only maximally consistent subset at both $w$ and $u$, so $\gcs{G}{w} = \cgcs{G}{w}$ and $\gcs{G}{u} = \cgcs{G}{u}$. Then, $wR_au$ and $wR_av$ for every $a \in G$, which by Euclidicity implies $uR_av$ for all such $a$. Thus, $v \in \cgcs{G}{u}$ and hence $u\crel{G}v$, as required by Euclidicity.
  \end{enumerate}
\end{enumerate}

\subsection*{Proof of \autoref{pro:DE:rel-properties}}\label{pro:DE:rel-properties:proof}

\begin{figure}[t]
  \begin{ctabular}{ccc}
    \begin{subfigure}[b]{0.24\textwidth}
      \centering
      \begin{tikzpicture}[frame rectangle, node distance = 2.5em and 1.5em, ->]
        \node[world] (w1) {w_1};
        \node[world, right = of w1] (w2) {w_2};
        \path (w1) edge [loop above] node [above = 2pt, label edge] {$b$} ()
                   edge node [above, label edge] {$a$} (w2)
              (w2) edge [loop above] node [above = 2pt, label edge] {$a,b$} ();
      \end{tikzpicture}
      \caption{$F_1$}
      \label{fig:f21}
    \end{subfigure}
    &
    \begin{subfigure}[b]{0.25\textwidth}
      \centering
      \begin{tikzpicture}[frame rectangle, node distance = 2.5em and 1.5em, ->]
        \node[world] (w) {w_1};
        \node[world, right = of w1] (w2) {w_2};
        \path (w) edge [loop above] node [above = 2pt, label edge] {$b$} ()
                   edge node [above, label edge] {$a$} (w2)
              (w2) edge [loop above] node [above = 2pt, label edge] {$b$} ()
                   edge (w);
      \end{tikzpicture}
      \caption{$F_2$}
      \label{fig:f22}
    \end{subfigure}
    &
    \begin{subfigure}[b]{0.26\textwidth}
      \centering
      \begin{tikzpicture}[frame rectangle, node distance = 3.5em and 1em, ->]
        \node[world] (w) {w_1};
        \node[world, below left = of w1] (w3) {w_3};
        \node[world, below  = of w1] (w2) {w_2};
        \node[world, below right = of w1] (w4) {w_4};
        \path (w) edge node [above = 2pt, label edge] {$a$} (w3)
                   edge node [within, label edge] {$a,b$} (w2)
                   edge node [above = 2pt, label edge] {$b$} (w4)
              (w2) edge node [above, label edge] {$a$} (w3)
                   edge node [above, label edge] {$b$} (w4)
              (w3) edge [loop above] node [above = 2pt, label edge] {$a,b$} ()
              (w4) edge [loop above] node [above = 2pt, label edge] {$a,b$} ();
      \end{tikzpicture}
      \caption{$F_3$}
      \label{fig:f23}
    \end{subfigure}
    \\[1em]
    \begin{subfigure}[b]{0.29\textwidth}
      \centering
      \begin{tikzpicture}[frame rectangle, node distance = 2.5em and 1.5em, ->]
        \node[world] (w) {w_1};
        \node[world, left = of w] (u1) {w_2};
        \node[world, right = of w] (u2) {w_3};
        \path (w)
            edge[<->] node [above, label edge] {$a$} (u1)
            edge[<->] node [above, label edge] {$b$} (u2)
          (u1) edge [loop above] node [above = 2pt, label edge] {$a,b$} ()
          (u2) edge [loop above] node [above = 2pt, label edge] {$a,b$} ();
      \end{tikzpicture}
      \caption{$F_4$}
      \label{fig:f24}
    \end{subfigure}
    &
    \begin{subfigure}[b]{0.28\textwidth}
      \centering
      \begin{tikzpicture}[frame rectangle, node distance = 2.5em and 1.5em, ->]
        \node[world] (w) {w_2};
        \node[world, left = of w] (u1) {w_1};
        \node[world, right = of w] (u2) {w_3};
        \path (w)
            edge[<->] node [above, label edge] {$a$} (u1)
            edge[<->] node [above, label edge] {$b$} (u2)
            edge [loop above] node [above = 2pt, label edge] {$a,b$} ()
          (u1) edge [loop above] node [above = 2pt, label edge] {$a$} ()
          (u2) edge [loop above] node [above = 2pt, label edge] {$b$} ();
      \end{tikzpicture}
      \caption{$F_5$}
      \label{fig:f25}
    \end{subfigure}
    &
    \begin{subfigure}[b]{0.3\textwidth}
      \centering
      \begin{tikzpicture}[frame rectangle, node distance = 1em and 1em, ->]
        \node[world] (w) {w_1};
        \node[world, below right = of w1] (w3) {w_3};
        \node[world, below left  = of w1] (w2) {w_2};
        \node[world, below right = of w2] (w4) {w_4};
        \path (w)  edge node [pos = 0.65, above = 2pt, label edge] {$b$} (w3)
                   edge node [pos = 0.65, above = 2pt, label edge] {$a$} (w2)
              (w2) edge node [above, label edge] {$b$} (w4)
                   edge [loop above] node [above, label edge] {$a$} ()
              (w3) edge [loop above] node [above, label edge] {$b$} ()
              (w3) edge node [above = 2pt, label edge] {$a$} (w4)
              (w4) edge [loop right] node [right = 2pt, label edge] {$a,b$} ();
      \end{tikzpicture}
      \caption{$F_6$}
      \label{fig:f26}
    \end{subfigure}
  \end{ctabular}
  \caption{Counterexamples used in the proof of \autoref{pro:DE:rel-properties}}
  \label{tbl:counterexamples-Npreservation}
\end{figure}
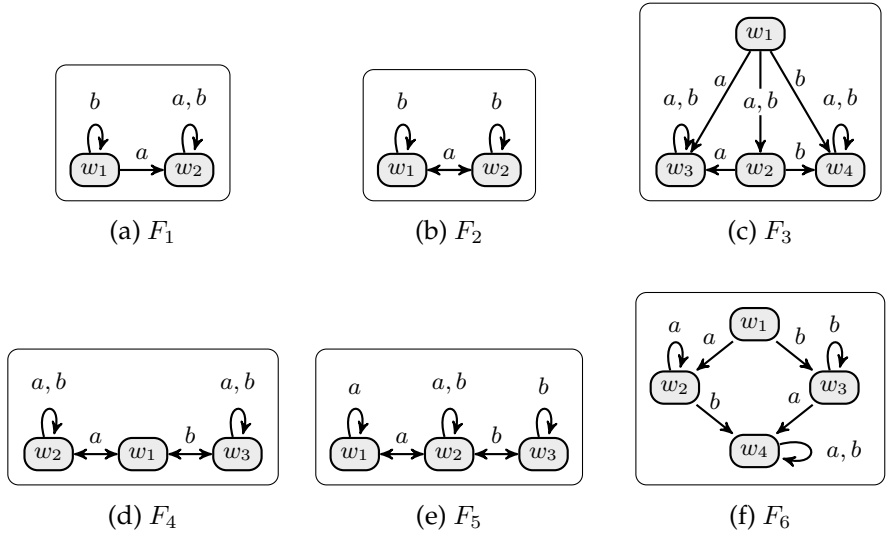

For the positive results, let $\M = \langle W,R,\val \rangle$ be a belief model.
\begin{enumerate}
  \item
  \begin{enumerate}[leftmargin=1.5em]
    \item\label{pro:DE:rel-properties:ser-ref} Whenever the individual relations are reflexive, $\gcs{G}{w}\neq \emptyset$ and $N_G(w) =\set{U \subseteq W \mid \gcs{G}{w} \subseteq U}$ for any $w \in W$. Thus, in reflexive models, it is never the case that $U, \comp{U} \in N_G(w)$.
    
    \item In frame $F_1$ \autorefp{fig:f21}, relations $R_a$ and $R_b$ are serial, so $N_a$ and $N_b$ satisfy $l_N$. They are also transitive and Euclidean. However, $\maxatin{\set{a}}{w_1}{\set{a,b}}$, $\maxatin{\set{b}}{w_1}{\set{a,b}}$ and $\comp{\gcs{\set{a}}{w_1}} =\set{w_1} = \gcs{\set{b}}{w_1}$, thus $N_{\set{a,b}}$ does not satisfy $l_N$.
    
    \item In frame $F_2$ \autorefp{fig:f22}, relations $R_a$ and $R_b$ are symmetric and serial. The conjecture sets at $w_1$ are the same as in the previous item.
    
    \item Seriality, symmetry and either transitivity or Euclidicity imply reflexivity, so \autoref{pro:DE:rel-properties:ser-ref} gives us the result.
  \end{enumerate}

  \item Whenever $r_N$ is satisfied for all agents (i.e., reflexive models), $w \in \gcs{G}{w}$ and thus $N_G(w)$ becomes $\set{U \subseteq W \mid \gcs{G}{w}\subseteq U}$, which clearly has the required property.

  \item
  \begin{enumerate}[leftmargin=1.5em]
    \item In the frame $F_3$ (\autoref{fig:f23}), relations $R_a$ and $R_b$ are serial and transitive (the latter also meaning that $t_N$ is satisfied for $N_a$ for all agents). In the frame, $\set{w_2} \in N_{\set{a,b}}(w_1)$. However, $\set{s \in F_3 \mid \set{w_2} \in N_{\set{a,b}}(s)}=\set{w_1}$, and $\set{w_1}\notin N_{\set{a,b}}(w_1)$.

    \item Assume $\M$ is reflexive and that $t_N$ is satisfied for the individual agents (and thus that the model is transitive). By reflexivity, for any $w \in W$ there is $u \in \gcs{G}{w}$ (the full group is consistent). When $t_N$ holds for $N_a$, $u \in \gcs{a}{w}$ implies $\gcs{a}{u} \subseteq \gcs{a}{w}$. Since this is the case for all $a \in G$, $\gcs{G}{u} \subseteq \gcs{G}{w}$. Thus $\forall u \in \gcs{G}{w}$, by superset closure of $N_G$ at $u$, $u \in \set{v \in W \mid \gcs{G}{w} \in N_G(v)}$ and for the same reason $u \in \set{v \in W \mid U \in N_G(v)}$ for arbitrary $U \supseteq \gcs{G}{w}$. By superset closure (of $N_G$ at $w$), $\set{v \in W \mid U \in N_G(v)} \in N_G(w)$ for any $U \in N_G(w)$.

    \item Assume $\M$ is symmetric and that $t_N$ is satisfied for the individual agents. Assume that $U \in N_G(w)$. Then $\exists \maxatin{G'}{w}{G}$ and $\gcs{G'}{w}\subseteq U$. Then, for arbitrary $u \in \gcs{G'}{w}$, $\gcs{G'}{u}\neq \emptyset$ (by symmetry of $R_a$ $\forall a \in G'$), $\gcs{G'}{u} \subseteq \gcs{G}{w}$ (by the transitivity of $R_a$ $\forall a \in G'$), and if $\exists H$ such that $G\subset \maxatin{H}{u}{G}$, then $\gcs{H}{u} \subseteq \gcs{G'}{u}$. Thus, by superset closure (at $u$), if $t\in \gcs{G}{w}$, then $u \in \set{v \in W \mid U \in N_G(v)}$, and by superset closure (at $w$), $\set{v \in W \mid U \in N_G(v)} \in N_G(w)$.

    \item Assume $\M$ is Euclidean, and that $t_N$ is satisfied for the individual agents. The argument is similar to the previous one: the only difference is that it is now the Euclidicity of the individual relations means that guarantee that if $u \in \gcs{G'}{w}$, then $\gcs{G'}{u} \neq \emptyset$.
  \end{enumerate}

  \item
  \begin{enumerate}[leftmargin=1.5em]
    \item In the frame $F_4$ (\autoref{fig:f24}), relations $R_a$ and $R_b$ are serial and  $N_a$ and $N_b$ satisfy $s_N$, and $w_1 \in \set{w_1}$. However $\comp{\set{w_1}} = \set{w_2,w_3}$, $\set{u \in F_4 \mid \set{w_2,w_3} \notin N_{\set{a,b}}(u)} = \emptyset$, and $\emptyset \notin N_{\set{a,b}}(w_1)$.

    \item Assume $\M$ is reflexive and that $t_N$ is satisfied for the individual agents (so $\M$ is symmetric). Assume that $w \in U \subseteq W$. It suffices to show that $\forall u \in \gcs{G}{w}$, $\comp{U} \notin \gcs{G}{u}$ (reflexivity guarantees $\gcs{G}{w} \neq \emptyset$, thus $\gcs{G}{w}\in N_G(w)$ and the by superset closure, the set containing also any remaining worlds that do not have the complement $U$ as a neighbourhood will be in $N_G(w)$). For any $u \in \gcs{G}{w}$, $\forall a \in G$: $R_awu$ (by definition of $\gcs{G}{w}$). By symmetry of the individual relations, $\forall a \in G$, $R_auw$. Thus, $w \in \gcs{G}{u}$. This means that for any $U' \in N_G(u)$, $w \in U'$, and thus $U' \neq \comp{U}$. \label{pro:DE:rel-properties:sym-ref}

    \item In the frame $F_5$ (\autoref{fig:f25}), relations $R_a$ and $R_b$ are transitive and Euclidean, and $N_a$ and $N_b$ satisfy $s_N$. It is also the case that $w_1 \in \set{w_1,w_3}$. However $\comp{\set{w_1,w_3}} = \set{w_2}$, $\set{s \in F_5 \mid \set{w_2} \notin N_{\set{a,b}}(s)} = \set{w_1,w_3}$, and $\set{w_1,w_3} \notin N_{\set{a,b}}(w_1)$.

    \item In any such class of models, if a model also satisfies $s_N$ for all individual agents, it is also reflexive (result given by \autoref{pro:DE:rel-properties:sym-ref}).
  \end{enumerate}

  \item
  \begin{enumerate}[leftmargin=1.5em]
    \item In the frame $F_6$ (\autoref{fig:f26}), relations $R_a$ and $R_b$ are transitive and serial, and $N_a$ and $N_b$ satisfy $e_N$. Now, $\set{w_4} \notin N_G(w_1)$. However, $\set{v \in W \mid \set{w_4} \notin N_G(v)}=\set{w_1}$, and $\set{w_1} \notin N_G(w_1)$.

    \item Assume $\M$ is reflexive, and that the neighbourhood functions of all agents satisfy $e_N$. Then, $\gcs{G}{w} \neq \emptyset$ (by reflexivity) and if $u \in \gcs{G}{w}$, then $\gcs{G}{w} \subseteq \gcs{G}{u}$ (by $e_N$/Euclidicity). This means that $N_G(u) \subseteq N_G(w)$, and thus if $U \notin N_G(w)$ then $\forall u \in \gcs{G}{w} \in N_G(w)$, $u \in \set{v \in W \mid U \notin N_G(v)}$, and by superset closure of $N_G$ at $w$, $\set{v \in W \mid U \notin N_G(v)} \in N_G(w)$. \label{pro:DE:rel-properties:euc-ref}

    \item In the frame $F_5$ (\autoref{fig:f25}), relations $R_a$ and $R_b$ are transitive and symmetric, and $N_a$ and $N_b$ satisfy $e_N$. Now, $\set{w_2} \notin N_G(w_1)$. However, $\set{v \in W \mid \set{w_2} \notin N_G(v)}=\set{w_1,w_3}$, and $\set{w_1,w_3} \notin N_G(w_1)$.
    
    \item In any such class of models, a model satisfying $s_N$ for all individual agents is also reflexive (result given by \autoref{pro:DE:rel-properties:euc-ref}).
  \end{enumerate}
\end{enumerate}

\subsection*{Proof of \autoref{pro:DE:invar}}\label{pro:DE:invar:proof}

Take arbitrary $\varphi \in \lang{\dbold{}{}}$. The proof proceeds by induction on the structure of $\varphi$, showing that if $\M,w \bidbold \M',w'$ then $\mps{w}\varphi$ iff $\M',w' \vDash \varphi$. The atomic case follows from the \textbf{atom} condition; the Boolean cases follow from the inductive hypotheses. The interesting case is $\dbold{G}{\psi}$.

{\prooflr} Assume that $\M,w \vDash \dbold{G}{\psi}$. Then, there is $U \subseteq \dom(\M)$ such that $w \colmcssleadto{G} U$ and $\forall u \in U$, $\M,u \vDash \psi$. By the \textbf{forth} condition, there is then some $U' \subseteq \dom(\M')$ such that $w' \colmcssleadto{G} U'$ and $U' \subseteq Z[U]$. By construction, for every $u' \in Z[U]$ there is some $u \in U$ such that $uZu'$. Thus, by inductive hypothesis, $\M',u' \vDash \psi$ holds for all $u' \in U' \subseteq Z[U]$. Then, $w' \colmcssleadto{G} U'$ (which means that $\exists\maxatin{H}{w'}{G}$ and $U' = \gcs{w}{H}$) implies $\M',w' \vDash \dbold{G}{\psi}$. 

{\proofrl} Analogous argument, using instead the \textbf{back} condition.

\subsection*{Proof of \autoref{pro:DE:implies}}\label{pro:DE:implies:proof}

We show that $Z = \set{(v,v') \in \dom(\M) \times \dom(\M') \mid v \eqdbold v'}$, the relation defined by $\eqdbold$, is a $\lang{\dbold{}{}}$-bisimulation. Take an arbitrary pair $(v,v')$ in it (so $v \eqdbold v'$); it will be show that the three conditions for $\lang{\dbold{}{}}$-bisimulation are satisfied.
\begin{bisimitemize}
  \item \textbf{Atom.} Immediate.
  
  \item \textbf{Forth.} Take $U \subseteq \dom(\M)$ such that $v \colmcssleadto{G} U$; for a contradiction, suppose there is no $U' \subseteq \dom (\M')$ such that $v' \colmcssleadto{G} U'$ and $U' \subseteq Z[U]$. Then, for all $U_i'$ such that $v' \colmcssleadto{G} U_i'$ we have $U_i' \not\subseteq Z[U]$, that is, there is $u_i' \in U_i'$ such that $uZu_i'$ fails for all $u \in U$. As $Z$ here is the $\eqdbold$-relation, the latter implies that every $U_i'$ such that $v' \colmcssleadto{G} U'_i$ contains a world $u'_i \in U_i'$ that can be distinguished from each $u_j \in U$ by some formula $\chi_j \in \lang{\dbold{}{}}$ (i.e., $\M,u_j \vDash \chi_j$ but $\M',u' \nvDash \chi_j$). Thus, the non-contradictory formula $\psi_i := \bigvee_{u_j \in U}\chi_j$ in $\lang{\dbold{}{}}$\footnote{The formula $\psi_i$ \begin{inlineenum} \item does not collapse to $\bot$ because $U$ is non-empty (it is the conjecture set of a maximally consistent group), and \item is a formula of $\lang{\dbold{}{}}$ because each $\chi_j$ is and, moreover, $U$ is finite (the image-finite assumption)\end{inlineenum}.} holds at every $u_j \in U$ (each $u_j$ satisfies its respective $\chi_j$) but fails at $u'_i \in U'_i$ (which falsifies all the disjuncts). Then, the non-tautological formula $\psi := \bigwedge_{v' \colmcssleadto{G} U'_i} \psi_i$ in $\lang{\dbold{}{}}$\footnote{The formula $\psi$ \begin{inlineenum} \item does not collapse to $\top$ because there is at least one $U' \subseteq \dom (\M')$ such that $v' \colmcssleadto{G} U'$ (otherwise, $\dbold{G}{\top}$ would distinguish $v$ and $v'$, contradicting the initial assumption $v \eqdbold v'$), and \item is a formula of $\lang{\dbold{}{}}$ because each $\chi_j$ is and, moreover, there is a finite number of $U'_i$ such that $v' \colmcssleadto{G} U_i'$ (the number of maximally consistent subgroups of a group is finite)\end{inlineenum}.} is such that:
  \begin{itemize}
    \item $\M,v \vDash \dbold{G}{\psi}$, as there is $U \subseteq \dom(\M)$ such that $v \colmcssleadto{G} U$ and, moreover, every world $u_j \in U$ satisfies every $\psi_i$, and hence satisfies $\psi$ too.
    
    \item $\M',v' \nvDash \dbold{G}{\psi}$, as every $U'_i \subseteq \dom(\M')$ that satisfies $v' \colmcssleadto{G} U'$ contains a world $u'_i$ that falsifies its corresponding $\psi_i$, and hence falsifies $\psi$ too.
  \end{itemize}
  Thus, $v$ and $v'$ can be distinguished by a formula in $\lang{\dbold{}{}}$, contradicting the initial $v \eqdbold v'$. Hence, there is some $U' \subseteq \dom(\M')$ such that $v' \colmcssleadto{G} U'$ and $U' \subseteq Z[U]$, and thus the \textbf{forth} condition holds for $Z$.

  \item \textbf{Back.} Analogous.
\end{bisimitemize}

\bibliographystyle{abbrvnat}
\bibliography{cdistbelbib}

\end{document}